\documentclass[11pt,a4paper]{article}

\usepackage[T1]{fontenc}
\usepackage[utf8]{inputenc}
\usepackage[a4paper,margin=2.45cm]{geometry}
\usepackage{amsmath}
\usepackage{booktabs}
\usepackage{longtable}
\usepackage{array}
\usepackage{graphicx}
\usepackage{newtxtext,newtxmath}
\usepackage{microtype}
\usepackage{setspace}
\usepackage[hidelinks,hypertexnames=false]{hyperref}

\hypersetup{
  pdftitle={Who Is Prioritized? Current Need versus Predicted Gains in China's Dibao Allocation: Evidence from a Revealed-Allocation Model},
  pdfauthor={Haojie Liu, Jiyuan Ling, and Zihan Lin}
}

\newcommand{\refentry}[1]{\par\noindent\hangindent=1.5em\hangafter=1 #1\par}

\begin{document}

\begin{center}
{\LARGE\bfseries Who Is Prioritized? Current Need versus Predicted Gains\\[0.25em]
in China's Dibao Allocation:\\[0.25em]
Evidence from a Revealed-Allocation Model}\\[1.1em]

Haojie Liu\textsuperscript{a,$\dagger$}\quad
Jiyuan Ling\textsuperscript{a,$\dagger$,*}\quad
Zihan Lin\textsuperscript{a,$\dagger$}\\[0.65em]

{\small
\textsuperscript{a}Department of Economics, University of California, Riverside\\
900 University Avenue, Riverside, CA 92521, United States\\[0.35em]
\textsuperscript{$\dagger$}These authors contributed equally to this work.\\
\textsuperscript{*}Corresponding author: Jiyuan Ling (\href{mailto:jling026@ucr.edu}{jling026@ucr.edu}).
}
\end{center}

\vspace{0.8em}

\begin{abstract}
Using the 2017--2021 China Household Finance Survey panel and a sample of 3,102 households that did not receive Dibao in 2017, this paper applies the BBK revealed-allocation model to examine how household economic circumstances and predicted policy gains explain observed Dibao receipt in China. We find that household income is the allocation criterion most difficult to substitute with other factors. When all gain-adjustment terms are positive, constraining the income-rank coefficient to be nonnegative reduces the model's original improvement in fit by approximately 33.6\%; by contrast, setting both predicted-gain coefficients for food consumption and education spending to zero still preserves approximately 95.6\% of that improvement. These results indicate that observed Dibao allocation places greater emphasis on prioritizing households facing current livelihood hardship than on pursuing predictable short-run improvements in expenditure outcomes. This allocation orientation reflects the trade-off in social assistance between basic-livelihood protection and expected policy gains and offers a new perspective on the potential long-run social-welfare value of basic-livelihood protection.
\end{abstract}

\vspace{0.45em}


\section{Introduction}

China's Minimum Living Standard Guarantee (Dibao) program uses household economic circumstances as central eligibility criteria, but the formal rules alone do not establish how much different household information contributes to actual benefit receipt. Using observed Dibao receipt, this paper asks two questions. When other household characteristics and predicted policy gains are considered together, can other factors replace the negative relationship between income and benefit receipt? How much additional information do predicted gains in food consumption and education spending provide for explaining actual receipt? These questions help clarify the role of household economic hardship in the program's observed allocation.

Poverty-targeting research has long considered how to identify disadvantaged households when resources and information are limited. Hanna and Olken (2018) discuss the trade-off between program coverage and transfer size; Brown et al. (2018) show that predicting poverty from observable household characteristics may leave some poor households unprotected; and Alatas et al. (2012) find that community assessments of poverty do not fully coincide with consumption-based measures. These studies show how poverty measures and available information influence who receives assistance. Using a cash transfer program in Kenya, Haushofer et al. (2025) compare recipient allocations based on initial deprivation and predicted policy impacts, extending poverty-targeting research to program objectives and social-welfare evaluation. Our study instead examines an existing Dibao program and asks how household conditions and predicted food and education gains explain observed benefit receipt.

Analyzing allocation criteria from policies already in place offers another approach to these questions. Bourguignon and Spadaro (2012) study welfare evaluations consistent with existing tax-and-transfer systems, and Hendren (2020) uses inverse-optimum weights to assess the efficiency of the prevailing tax system. Björkegren et al. (2026, hereafter BBK) incorporate household characteristics and predicted gains in multiple outcomes into a revealed-allocation model to examine which factors explain observed policy choices. Conventional receipt regressions estimate associations between household characteristics and benefit receipt. The BBK framework additionally allows us to constrain the coefficient on a characteristic, reoptimize other parameters, and measure the resulting loss of model fit. We use this approach to assess whether other characteristics and predicted gains can replace the negative relationship between income and Dibao receipt.

Dibao provides a concrete institutional setting for this analysis. The 2014 Interim Measures for Social Assistance establish basic-livelihood protection, social equity, and social harmony and stability as program objectives and make household income and assets central eligibility criteria (State Council, 2014). Actual receipt also depends on applications, economic verification, grassroots participation, and county-level approval, while local thresholds and implementation differ across regions. Prior studies examine urban and rural coverage, poverty targeting, and poverty reduction (Chen et al., 2006; Golan et al., 2017), as well as the effects of local thresholds and administrative discretion (Ravallion, 2009; Li and Walker, 2021). Han and Gao (2019) study elite capture in rural community-based targeting, while Ravallion (2008) discusses how poverty measurement changes the assessment of targeting. Building on this literature, we examine the importance of household income and other information in explaining observed Dibao receipt.

We use the 2017–2021 China Household Finance Survey (CHFS) panel to examine Dibao receipt in 2019 among households not receiving benefits in 2017. The structural sample contains 3,102 households, 265 of which received Dibao in 2019, across 125 county (or district)-by-urban/rural comparison sets in 107 counties. We use changes in food consumption and education spending between 2017 and 2021 to predict household-level policy gains. The BBK model combines seven household characteristics and the two predicted gains in an allocation score. We restrict the signs of selected household-characteristic coefficients and reoptimize the remaining parameters to measure the cost of replacing those relationships. We also vary the predicted-gain coefficients and simulate recipient lists under fixed quotas to examine changes in household rankings and simulated selections.

When the gain-adjustment term is positive for every household, constraining the income-rank coefficient to be nonnegative and reoptimizing other parameters sacrifices approximately 33.6\% of the original improvement in in-sample fit; the corresponding costs for household-head education and elderly household membership are 5.4\% and 4.6\%. Setting both predicted-gain coefficients to zero and reoptimizing the household-characteristic coefficients preserves approximately 95.6\% of the original fit improvement. Fixed-quota simulations also show that different parameter combinations can select similar shares of low-income households but different individual recipients.

These results indicate that household income plays a difficult-to-substitute role in explaining observed Dibao receipt and are consistent with the program's need-based role in protecting the basic livelihoods of disadvantaged households. Section 7 accordingly discusses the distinction between need-based targeting and targeting sustained poverty exit, and the conditions under which prioritizing the poorest households may be consistent with maximizing social welfare.

Section 2 introduces Dibao's institutional background and allocation objectives. Section 3 presents the model and estimation methods, Section 4 the data and predicted gains, and Sections 5 and 6 the empirical results. Section 7 discusses their economic and policy implications, and Section 8 concludes.

\section{Institutional Background and Allocation Objectives of Dibao}

\subsection{Program Objectives and Economic Eligibility}

China’s Dibao program provides basic-livelihood support to households facing economic hardship. The 2014 Interim Measures for Social Assistance identify “safeguarding citizens’ basic livelihoods, promoting social equity, and maintaining social harmony and stability” as objectives of social assistance, and require assistance to follow the principles of openness, fairness, impartiality, and timeliness. Household income, assets, and other economic circumstances are key criteria for Dibao eligibility (State Council, 2014).

During the paper’s 2017–2019 observation period, applicants had to meet local household-registration, income, and asset requirements. Under the 2012 Measures for Review and Approval of Minimum Living Standard Guarantee Applications (Trial), households with local permanent residence registration could apply if per capita income among family members living together was below the local Dibao threshold and their assets met the prescribed conditions. Eligibility assessments are conducted at the household level and verify co-resident family members, disposable income over a specified period, and household assets (Ministry of Civil Affairs, 2012). Household composition, income, and assets therefore provide the principal information used in determining economic eligibility.

Eligibility standards also vary across locations. The Interim Measures provide that provincial governments or municipal governments with districts set Dibao thresholds based on local costs of basic necessities and adjust them as economic and social conditions and prices change. Corresponding local governments also formulate income and asset assessment rules in accordance with national requirements. Consequently, identical household income may lead to different eligibility judgments in different locations. Ravallion (2009) also examines the relationship between local eligibility thresholds and economic conditions in urban China. These differences provide an important institutional context for understanding observed Dibao receipt.

\subsection{Household Application, Economic Verification, and Local Implementation}

During the period studied, Dibao eligibility determination generally began when a household applied for assistance. Households submitted a written application to the township government or subdistrict office where their household registration was held, reported household income and assets, and authorized verification of their economic circumstances. Those unable to apply themselves could ask a village or residents’ committee to submit an application on their behalf. With support from these committees, township governments or subdistrict offices verified household economic conditions and living circumstances through information checks, home visits, and interviews with neighbors, and organized community appraisals. After public disclosure of the review findings, materials were forwarded to the county civil affairs department for examination and a final decision; rejected applicants received written reasons (Ministry of Civil Affairs, 2012). The process linked household reporting, grassroots verification, and county-level approval.

Grassroots organizations could also supply information beyond the household’s application. Employment status, family support obligations, and everyday hardship could be supplemented and checked through home visits and community appraisals. Kuhn et al. (2016) find that in some rural areas, funding or recipient-slot limits could lead village organizations to screen applications before formal submission. Grassroots organizations therefore helped verify applicants’ circumstances and could influence which households proceeded to later review, while the county civil affairs department retained final authority over eligibility decisions.

Dibao eligibility was subject to ongoing review. Under the Interim Measures for Social Assistance, relevant authorities periodically checked recipient households’ composition, income, and assets, adjusting or discontinuing transfers when circumstances changed. Receipt at any survey date consequently reflected not only earlier applications and approvals but also subsequent changes in household economic conditions and eligibility reassessments.

\subsection{Changes in Observed Dibao Receipt and the Scope of Comparison}

Using CHFS records of Dibao receipt for the same households in 2017 and 2019, we define an observed-entry indicator. Households that did not receive Dibao in 2017 but received it in 2019 are classified as entrants; households receiving Dibao in neither wave are classified as nonentrants. Starting from the common group of nonrecipients in 2017, we examine receipt in 2019 and use this indicator as the observed outcome in the subsequent receipt regressions and allocation analysis.

Because Dibao eligibility standards and implementation differ by locality, we use verifiable county (or district) and urban–rural identifiers in CHFS to place households from the same county (or district) and urban–rural category in an empirical comparison unit. The BBK model compares household rankings within these units, and fixed-quota simulations are also conducted at the unit level. This places household comparisons within broadly similar local institutional conditions. Sections 3 and 4 describe sample selection, construction of the comparison units, and estimation methods.

\section{Theoretical Framework and Empirical Method}

\subsection{Policy Value, Household Component, and Predicted Gains}

We use the revealed-allocation framework of Björkegren et al. (2026, hereafter BBK) to incorporate household circumstances and potential improvements from assistance into a common measure of policy value. Let $T_i\in\{0,1\}$ indicate whether household $i$ receives assistance, and let $v_{ij}(T_i)$ denote outcome $j$ under the corresponding state. BBK express household policy value as

\begin{equation}
\begin{aligned}
S_i(T_i) &= w_i u_i(T_i),\\
u_i(T_i) &= \sum_j b_{ij}v_{ij}(T_i)+a_iT_i.
\end{aligned}
\tag{3.1}\label{eq:3-1}
\end{equation}

Here $w_i$ is the model’s household evaluation weight, $b_{ij}$ the valuation coefficient for outcome $j$, and $a_i$ the baseline value of providing assistance. Comparing the household’s policy value with and without assistance, define $\Delta v_{ij}=v_{ij}(1)-v_{ij}(0)$. This gives the incremental policy value from assistance:

\begin{equation}
\Delta S_i
=w_i\left[a_i+\sum_j b_{ij}\Delta v_{ij}\right]
=w_i a_i\left[1+\sum_j\left(\frac{b_{ij}}{a_i}\right)\Delta v_{ij}\right],
\qquad a_i>0.
\tag{3.2}\label{eq:3-2}
\end{equation}

Equation~\eqref{eq:3-2} links the baseline value of assistance to improvements in different outcomes. We combine $w_i a_i$ into the household component $\omega_i$ and assume that each outcome dimension has the same coefficient relative to baseline assistance value across households:

\begin{equation}
\begin{aligned}
\frac{b_{ij}}{a_i} &= \beta_j \qquad \text{for all households } i,\\
\Delta S_i &= \omega_i\left[1+\sum_j\beta_j\Delta v_{ij}\right].
\end{aligned}
\tag{3.3}\label{eq:3-3}
\end{equation}

The incremental policy value thus has two components: the household component $\omega_i$ captures differences in baseline value, while the gain adjustment in parentheses captures improvements in outcomes and their coefficients. Two households with the same predicted gains may have different household components; similarly, differences in predicted gains may change their allocation order even when the household components are similar.

In the empirical analysis, we consider food consumption and education spending and use fixed conditional gain predictions $\hat\tau_{iC}$ and $\hat\tau_{iE}$. The vector $X_i$ contains seven household characteristics: income rank, household size, indicators for children and elderly members, a female household head, household-head age, and household-head education. We model the household component as an exponential function without an intercept and construct the complete allocation score as follows:

\begin{equation}
\begin{aligned}
\omega(X_i;\theta) &= \exp(X_i'\theta),\\
G_i(\beta) &= 1+\beta_C\widehat{\tau}_{iC}+\beta_E\widehat{\tau}_{iE},\\
Q_i(\theta,\beta) &= \omega(X_i;\theta)G_i(\beta),
\qquad \beta_C,\beta_E\geq 0.
\end{aligned}
\tag{3.4}\label{eq:3-4}
\end{equation}

Equation~\eqref{eq:3-4} shows that the seven household characteristics enter the score through $\omega(X_i;\theta)$, while predicted food and education gains adjust the score through $G_i(\beta)$. The vector $\theta$ describes conditional relationships between characteristics and the household component: with other characteristics fixed, positive coefficients correspond to a higher household component and negative coefficients to a lower one. The coefficients $\beta_C$ and $\beta_E$ describe the roles of the predicted gains in the score; their numerical interpretation depends on the outcome scales and the baseline term of one. Section 4 defines the variables.

The complete allocation score is the product of the household component and the gain adjustment. When $G_i(\beta)>0$, holding predicted gains and coefficients fixed, a higher household component yields a higher complete score. We therefore interpret the directions of household-characteristic coefficients using parameter solutions that satisfy this positivity condition for every sample household. Appendix E reports the condition and checks.

\subsection{Observed Entry and Allocation Fit}

Let $D_i$ denote the observed Dibao entry indicator defined in Section 2, and let $m$ index the comparison sets formed by county (or district) and urban or rural status. Using the allocation score $Q_i$ from Section 3.1, the model compares households within each set and fits observed entries across sets with a common parameter vector. Survey weights $s_i$ are normalized to have mean one over the entire structural sample, preserving their relative sizes across households and sets.

To relate allocation scores to observed receipt, we define $W_m$ as the total survey weight of entrants in set $m$, $q_i$ as household $i$’s share of entrant weight, and $p_i$ as the softmax fitted share based on the complete allocation score:

\begin{equation}
W_m=\sum_{i\in m}s_iD_i,
\qquad
q_i=\frac{s_iD_i}{W_m},
\qquad
p_i(Q)=\frac{\exp(Q_i)}{\sum_{k\in m}\exp(Q_k)}.
\tag{3.5}\label{eq:3-5}
\end{equation}

The term $q_i$ describes the weighted distribution of actual entrants within each set; $q_i$ equals zero for nonentrants. The fitted share $p_i$ is computed from the scores of all households in the set, with shares summing to one. Higher-scoring households receive larger fitted shares. We thus assess whether actual entrants receive comparatively high allocation scores and fitted shares.

We estimate the parameters by minimizing this weighted loss:

\begin{equation}
\begin{aligned}
L_m(\theta,\beta)
&=-W_m\sum_{i\in m}q_i\log p_i(Q),\\
L(\theta,\beta)
&=\sum_m L_m(\theta,\beta)
=-\sum_m\sum_{i\in m}s_iD_i\log p_i(Q).
\end{aligned}
\tag{3.6}\label{eq:3-6}
\end{equation}

This loss measures the fitted shares assigned to actual entrants. Entrants enter directly with survey weights; nonentrants affect the shares through the softmax denominator. Multiple entrants within a set share that denominator, jointly determining the entrant distribution.

The aggregate loss sums the set-level losses while retaining differences in total entrant weight $W_m$. Sets with greater entrant weight therefore contribute more to the objective. Estimation seeks parameter values that minimize this aggregate loss and fit observed Dibao entry across locations. Appendix B details the score scale, softmax temperature, objective derivatives, and numerical implementation.

\subsection{Distinguishing Household Conditions from Policy Gains}

The household component and predicted gains draw on the same observed information. The seven household characteristics directly determine the household component and are also used to predict gains in food consumption and education spending. Appendix J verifies that both gain predictions can be represented linearly by these seven characteristics and a constant. Separating the two components' parameters therefore relies primarily on functional form: the household component is exponential, the gain adjustment is linear, and their product determines the complete allocation score. Parameter restrictions also affect the scores and rankings different parameter combinations can produce.

The shared information allows the two components to compensate for one another. When every household's gain adjustment is strictly positive, we can take logarithms of the complete allocation score:

\begin{equation}
\log Q_i(\theta,\beta)=X_i'\theta+\log G_i(\beta),\qquad G_i(\beta)>0.
\tag{3.7}\label{eq:3-7}
\end{equation}

If the household and gain coefficients change from $(\theta,\beta)$ to $(\theta',\beta')$, with strictly positive gain adjustments both before and after, the change in household $i$'s log score is

\begin{equation}
\log Q_i(\theta',\beta')-\log Q_i(\theta,\beta)=X_i'(\theta'-\theta)+\log\left(\frac{G_i(\beta')}{G_i(\beta)}\right).
\tag{3.8}\label{eq:3-8}
\end{equation}

This expression shows how compensation can arise. Changing the food or education gain coefficients changes the gain adjustment, while reoptimizing the household-characteristic coefficients may offset the resulting score changes. The complete scores and rankings can therefore remain similar. To assess the additional information provided by predicted gains, we examine how much of the original fit the model retains after reoptimizing the household component.

We study this compensation using 57 recorded solutions in the specified 1\% raw-loss-tolerance region that satisfy strict positivity. For each alternative gain coefficient vector $\beta^{(r)}$, we calculate two complete allocation scores:

\begin{equation}
Q_i^{\mathrm{fixed},r}=\omega(X_i;\hat\theta^{(0)})G_i(\beta^{(r)}),
\tag{3.9}\label{eq:3-9}
\end{equation}

\begin{equation}
Q_i^{\mathrm{reopt},r}=\omega(X_i;\hat\theta^{(r)})G_i(\beta^{(r)}).
\tag{3.10}\label{eq:3-10}
\end{equation}

Here $\hat\theta^{(0)}$ denotes the baseline household-component coefficients and $\hat\theta^{(r)}$ the household-component coefficients reoptimized for the alternative gain vector. The first calculation holds the baseline household component fixed and changes only the gain coefficients, isolating their effect on household rankings. The second uses the reoptimized household component and compares rankings after both components adjust. Comparing the two indicates how far reoptimization offsets the ranking changes induced by the gain coefficients. Appendices E and H describe the 57-solution selection, strict-positivity requirement, and ranking calculations.

The directional constraints and parameter profiles below follow the same logic: fix or restrict selected parameters, reoptimize the rest, and compare the fitted loss. This measures how much original fit the other parameters can retain when a household relationship or gain coefficient changes.

\subsection{Comparing Alternative Explanations of Allocation}

We compare the explanatory roles of household characteristics and predicted gains by restricting selected model parameters, reoptimizing the others, and examining the resulting fit loss. If the remaining parameters can preserve almost the original fit, the restricted relationship is relatively easy to replace. If the loss remains much higher after reoptimization, the relationship is difficult to replace. Throughout, we hold the household sample, variable definitions, and predicted food and education gains fixed and use the same allocation loss.

For household characteristics, we separately constrain the household-component coefficients on income rank and household-head education to be nonnegative and the coefficient on elderly-member status to be nonpositive. In the baseline model, the income-rank coefficient is negative: higher income rank corresponds to a smaller household component, holding other characteristics fixed. With the coefficient constrained to be nonnegative, we reoptimize the other six household-characteristic coefficients and both predicted-gain coefficients and record the lowest loss attainable under this restriction. The loss increase measures how far other parameters can substitute for the negative income relationship. We apply the same method to education and elderly-member status.

For predicted policy gains, we first jointly set the food-consumption and education-spending gain coefficients to zero, $\beta_C=\beta_E=0$, and reoptimize all seven household-component coefficients. The complete allocation score then reduces to the household component:

\begin{equation}
Q_i(\theta,0)=\exp(X_i'\theta).
\tag{3.11}\label{eq:3-11}
\end{equation}

Comparing the zero-gain model's lowest recorded loss with the full model's loss measures the additional in-sample fit provided jointly by the two predicted gains when the household coefficients can adjust. On the existing scales of the food and education outcomes, we also impose $\beta_E\geq\beta_C+0.05$ and reoptimize the remaining parameters to measure the fit cost of reversing the relative order of the two gain coefficients.

We then examine how well alternative gain-coefficient values can sustain the original fit through parameter profiles. For a given gain vector $\beta=(\beta_C,\beta_E)$, define the profile loss as

\begin{equation}
L_p(\beta)=\min_{\theta\in\Theta}L(\theta,\beta).
\tag{3.12}\label{eq:3-12}
\end{equation}

Here $\Theta$ is the prespecified range of household-component parameters. The joint profile holds both food and education coefficients fixed while reoptimizing the seven household coefficients; a one-dimensional profile fixes one gain coefficient while optimizing the other gain coefficient and the household coefficients. Recording the lowest loss at every candidate coordinate shows which gain values and combinations maintain a fit near the baseline. Appendices B--D report the parameter-search ranges, evaluated coordinates, and specified loss-tolerance thresholds.

For a common scale across restriction tests, we use the full model's fit improvement relative to an equal-score model as the denominator. Let $L^*$ be the lowest recorded loss from the baseline search. When all households within comparison set $m$ have equal scores, $p_i=1/N_m$, where $N_m$ is the number of households in that set. Substitution into Equation~\eqref{eq:3-6} gives the equal-score loss

\begin{equation}
L^{\mathrm{eq}}=\sum_m W_m\log N_m.
\tag{3.13}\label{eq:3-13}
\end{equation}

The full model thus improves on the equal-score benchmark by $L^{\mathrm{eq}}-L^*$. An alternative model with loss $L$ raises the loss relative to the full model by $L-L^*$. Dividing the latter by the former gives the fraction of original fit improvement forfeited:

\begin{equation}
r(L)=\frac{L-L^*}{L^{\mathrm{eq}}-L^*}.
\tag{3.14}\label{eq:fit-sacrifice}
\end{equation}

This measure quantifies how much original fit an alternative model gives up. $r(L)=0$ corresponds to the baseline fit; $r(L)=1$ corresponds to the equal-score benchmark. The fraction retained is $1-r(L)$. For instance, $r(L)=0.05$ means that an alternative forfeits 5\% of the original fit improvement and retains 95\%. We apply this measure to household-direction restrictions, the zero-gain model, and the gain-order restriction.

The parameter profiles also identify specified loss-tolerance regions using the proportional increase in original total loss, which has denominator $L^*$ rather than $L^{\mathrm{eq}}-L^*$ in Equation~\eqref{eq:fit-sacrifice}. These two measures respectively describe relative total-loss increases and the fraction of original fit improvement forfeited. Appendix B gives their conversion. Directional-constraint and gain-order searches require $G_i(\beta)>0$ for every household, whereas baseline and prespecified sensitivity estimates check positivity after optimization.

Finally, we test sensitivity to the softmax temperature, comparison-set weights, and fit loss. County-level reweighting uses the same 30 pregenerated county multipliers in five specifications, changing counties' relative contributions to total loss. Both the sample and gain-prediction inputs remain fixed. Appendix G reports all specifications, weight construction, optimization procedures, and numerical diagnostics.

\subsection{Household Rankings and Simulated Lists}

Even when different parameter combinations achieve similar model fit, they may change the relative rankings of households and the specific households selected. We use the complete allocation score $Q_i$ to rank households from highest to lowest within each county (or district) by urban/rural comparison set, fixing the simulated number of recipients at the observed number of Dibao entrants in that set:

\begin{equation}
K_m=\sum_{i\in m}D_i.
\tag{3.15}\label{eq:3-15}
\end{equation}

For each parameter combination, we select the $K_m$ highest-scoring households as simulated recipients. Any score ties are resolved using a fixed household identifier. Since both the households and the number of simulated places in each set remain unchanged, differences between lists represent substitutions caused by changes in rankings. The households observed entering Dibao in 2019 provide a reference list.

We compare allocation outcomes across parameter combinations along three dimensions. First, correlations between complete-score rankings show whether the relative positions of all households change appreciably. Second, we compare overlap between simulated lists and between a simulated list and observed entrants. For two lists $A$ and $B$ with the same set-level quotas, overall overlap is

\begin{equation}
\operatorname{Overlap}(A,B)=\frac{\sum_m |A_m\cap B_m|}{\sum_m K_m}.
\tag{3.16}\label{eq:3-16}
\end{equation}

Here $A_m$ and $B_m$ are the households selected by the two lists in comparison set $m$. The numerator counts households selected by both lists; the denominator is the total number of simulated places across sets. We calculate overlap between alternative and baseline lists, pairwise overlap across alternative lists, and the fraction of households on the baseline list that remain selected in all comparison solutions. These statistics describe different aspects of the stability of individual selections.

Finally, we examine the composition of simulated lists and individual selection outcomes. List shares for income groups and households with elderly members show whether different parameter combinations change the overall composition of recipients. For individual households, we record selection status across parameter combinations and relate changes to a household's position relative to the baseline selection cutoff. Appendices G, H, and L provide the details on ranking statistics, overlap denominators, household classifications, and distances from the selection cutoff.

Complete rankings and simulated lists capture different dimensions of allocation change. Complete rankings use the relative positions of all households within a comparison set, whereas simulated lists depend only on whether households fall within the top $K_m$ places. Even a small change in overall rankings may swap households around the selection cutoff, changing who is selected. We therefore report complete rankings, recipient-group composition, and household-level lists together.

\section{Data and Predicted Policy Gains}

\subsection{CHFS Panel and Analysis Samples}

We use the 2017--2021 household panel of the China Household Finance Survey (CHFS), which collects information on household income, assets, consumption spending, demographic characteristics, and participation in social protection programs (Gan et al., 2014). We study households not receiving Dibao in 2017 and classify them as entrants or nonentrants according to their receipt status in 2019. This status is the observed outcome in our subsequent allocation analysis. Changes in food consumption and education spending from 2017 to 2021 are used to predict potential improvements associated with receiving Dibao. Thus, we use 2019 receipt records to study observed selection and the 2017--2021 spending changes to estimate policy gains.

Sample selection starts with an existing analysis frame of 6,651 households. Checks of longitudinal household links, exclusion of households receiving Dibao in 2017, and the prespecified common-support restriction leave a prediction population of 6,132 households. All have complete model covariates and positive survey weights. Food consumption is observed for 5,724 households and education spending for 5,768 households; these samples are used to estimate the respective policy gains. We select the structural analysis sample from the 6,132-household prediction population by defining comparison sets using county (or district) and urban/rural status, then retaining sets with at least three households and both entrants and nonentrants. The final structural sample contains 3,102 households in 125 comparison sets across 107 counties, including 265 households receiving Dibao in 2019.

The food, education, and structural analysis samples are three samples drawn from the same prediction population. Estimating the food and education gains requires observed values of the corresponding outcomes. The structural analysis instead requires household characteristics, Dibao receipt records, and predictions of both gains. Because the first-stage models can generate predictions from household characteristics, some households with missing realized food or education outcomes can still be included in the structural sample. We thus use observed spending changes to estimate the gains and then use the predictions to explain observed Dibao receipt. Appendix A.1 documents sample selection, entrant counts at each stage, and outcome missingness.

\subsection{Outcome Dimensions and Household Characteristics}

We select food consumption and education spending as the two outcomes for predicting policy gains. Food consumption directly captures basic household consumption, while education spending reflects a household's investment in education and training. These dimensions allow us to examine potential improvements in basic consumption and changes in educational spending associated with Dibao.

The food outcome is the change in log real food consumption per capita from 2017 to 2021. We first divide nominal household food spending by household size in each year and then deflate it using the relevant consumer price index (CPI). Let $C_{it}$ denote household $i$'s nominal food expenditure in year $t$, $n_{it}$ household size, and $d_t$ the price deflator. Real per capita food consumption and its change are

\begin{equation}
C^{\mathrm{real,pc}}_{it}=\frac{C_{it}}{n_{it}d_t},
\qquad
\Delta Y_i^C
=\log C^{\mathrm{real,pc}}_{i,2021}
-\log C^{\mathrm{real,pc}}_{i,2017}.
\tag{4.1}\label{eq:4-1}
\end{equation}

The education outcome uses real total household spending on education and training. We deflate spending in each year, apply the $\log(1+x)$ transformation, and calculate the difference between 2017 and 2021. This transformation retains households with zero education spending. Unlike the per capita food measure, education spending is measured at the household level. The two outcomes capture changes in real food consumption per capita and real household education spending, and their numerical scales reflect their respective units and transformations.

We use seven household characteristics to construct the household component and predict food and education gains: income rank, household size, an indicator for children, an indicator for elderly members, a female household head, household-head age, and household-head education. Income rank measures a household's relative position in total household income; household size and the child and elderly indicators describe household composition; and the head's sex, age, and education describe the household head. Using the same seven characteristics in both components allows us to compare their contributions to explaining observed Dibao receipt on a common information basis.

Income rank is calculated using survey weights. The child indicator identifies households with members aged 0--15; the elderly indicator identifies members aged 60 or older. Household-head education follows the survey's ordered 1--9 coding. Urban/rural status enters propensity-score estimation, outcome nuisance models, and the construction of comparison sets, but not the heterogeneous-gain function as a separate covariate. Appendices A.2--A.4 document variable coding, outcome scales, and income-rank construction.

\subsection{First-Stage Prediction and Inputs to the Structural Model}

We first use household characteristics to predict the outcome gains households may obtain from Dibao, then enter these predictions into the BBK allocation model. For food consumption and education spending, the policy gain is the conditional mean difference between potential outcome changes under entry and nonentry, given household characteristics.

Let $\widetilde X_i$ denote the first-stage household characteristics. For outcome $j$, define

\begin{equation}
\tau_j(\widetilde X_i)
=E\!\left[v_{ij}(1)-v_{ij}(0)\mid\widetilde X_i\right],
\qquad
\widehat\tau_{ij}=\widehat\tau_j(\widetilde X_i).
\tag{4.2}\label{eq:4-2}
\end{equation}

Here $\tau_j(\widetilde X_i)$ is the conditional mean policy gain, and $\widehat{\tau}_{ij}$ is the first-stage prediction for household $i$. A causal interpretation requires three conditions: conditional on first-stage covariates, entry status is independent of potential outcome changes; both entry states have positive conditional probabilities; and observed outcomes agree with the potential outcomes under households' realized entry states (Nie and Wager, 2021).

The structural model uses food and education gain predictions from the implemented household-level five-fold cross-fitted procedure. This procedure forms five internal folds at the household level and cross-fits the propensity score and outcome nuisance models. Using residuals from the samples with observed food and education outcomes, it then fits separate linear R-learners with an intercept and the seven household characteristics. These models generate both predicted gains for each household in the 6,132-household prediction population. Related methods are described in Nie and Wager (2021) and Chernozhukov et al. (2018); Appendix A.4 documents the implementation.

We then hold the two predictions fixed as inputs to the structural model. Following the allocation-score equation in Section 3, each predicted gain is weighted by its corresponding coefficient and combined with the household component to construct the full score. All subsequent parameter comparisons use the same predicted-gain vectors, varying only the household-component parameters and gain coefficients.

The first stage thus estimates how predicted policy gains vary with household characteristics; the second examines how much additional information those predicted differences contribute to explaining observed Dibao entry.

\subsection{County-Held-Out Prediction Diagnostic}

To assess whether household-level differences in predicted gains carry over to other locations, we separately conduct a five-fold county-held-out diagnostic. Counties are assigned to training and test folds. Test counties in each fold are excluded from that fold's propensity-score, outcome-nuisance, and gain-function estimation. Within the training counties, nuisance models are still cross-fitted using household-level folds. The diagnostic therefore evaluates predicted gain heterogeneity in counties excluded from training.

For the diagnostic, the propensity-score and outcome nuisance models use logistic and ridge regressions, respectively. Nuisance-model selection in the implemented education-gain procedure also considers histogram gradient boosting. The county-held-out exercise consequently assesses a related but somewhat different prediction procedure; the structural allocation analysis continues to use the original fixed gain predictions.

We compare heterogeneous-gain predictions with a constant-effect benchmark estimated from the training folds and evaluated in the same test counties. Predictive performance is measured using R-loss: the weighted squared error from explaining the residualized outcome with the product of the predicted gain and residualized Dibao entry. Lower R-loss indicates better prediction.

We define improvement as

\begin{equation}
\text{Improvement}=R\text{-loss}_{\text{constant effect}}-R\text{-loss}_{\text{heterogeneous prediction}}.
\tag{4.3}\label{eq:4-3}
\end{equation}

Positive improvement indicates better performance from heterogeneous prediction; negative improvement indicates higher loss. Survey weights are applied within each test fold, and fold-level results are aggregated using test-fold household counts. Table~\ref{tab:county-heldout} reports the sample sizes and predictive results.

\begin{table}[htbp]
\centering
\caption{Analysis Samples and County-Held-Out Prediction Diagnostic}
\label{tab:county-heldout}
\scriptsize
\setlength{\tabcolsep}{3.5pt}
\begin{tabular}{lrrrrrr}
\toprule
Sample/outcome & Households & Entrants & \shortstack{Heterogeneous\\R-loss} & \shortstack{Constant-effect\\R-loss} & Improvement & \shortstack{No-intercept\\slope} \\
\midrule
Structural sample & 3,102 & 265 & --- & --- & --- & --- \\
Food & 5,724 & 252 & 0.774596 & 0.771547 & $-0.003049$ & $-0.085$ \\
Education & 5,768 & 248 & 18.083475 & 18.022391 & $-0.061084$ & $-0.362$ \\
\bottomrule
\end{tabular}

\vspace{0.35em}
\begin{minipage}{0.96\textwidth}
\footnotesize\textit{Note:} R-loss and the no-intercept projection coefficient are calculated within each test fold and then aggregated using test-fold household counts. Appendix A.4 defines the projection coefficient and reports fold-level results.
\end{minipage}
\end{table}

Table~\ref{tab:county-heldout} shows that average heterogeneous-prediction R-loss exceeds the constant-effect benchmark for both outcomes. The improvements are $-0.003049$ for food and $-0.061084$ for education; the corresponding no-intercept projection coefficients are $-0.085$ and $-0.362$. In this county-held-out diagnostic, using household characteristics to predict gain heterogeneity does not improve average predictive performance for household outcome changes in test counties.

These results indicate limited incremental information from the available household characteristics and prediction procedure for distinguishing gains outside the training counties. Section 6 examines the separate question of how much additional fit food and education predictions contribute to observed Dibao receipt, and how much of that contribution can be replaced by the household component.

\section{Household Economic Conditions and Observed Dibao Entry}

\subsection{The Negative Relationship between Income and Dibao Entry}

We begin with receipt regressions to examine the relationship between household economic circumstances and Dibao entry. The structural sample contains 3,102 households that did not receive Dibao in 2017, of which 265 received it in 2019. Table~\ref{tab:entry-regressions} reports three specifications: a Logit model, a linear probability model (LPM), and an LPM with comparison-set fixed effects. All three control for seven household characteristics and use survey weights and county-clustered standard errors.

The income-rank coefficients are $-2.2672$, $-0.1617$, and $-0.1644$, respectively. All are negative, indicating that households with lower relative income are more likely to enter Dibao, conditional on other household characteristics. After adding comparison-set fixed effects, the income coefficient remains similar in sign and magnitude to the ordinary LPM estimate. Because the comparison sets are defined by county (or district) and urban/rural status, this relationship is also evident among households within the same local comparison set.

\begin{table}[htbp]
\centering
\caption{Household Economic Conditions and Dibao Entry: Receipt Regressions}
\label{tab:entry-regressions}
\small
\begin{tabular}{lccc}
\toprule
Household characteristic & Logit & LPM & \shortstack{Comparison-set\\FE LPM} \\
\midrule
Income rank & $-2.2672$ & $-0.1617$ & $-0.1644$ \\
            & $(0.4136)$ & $(0.0265)$ & $(0.0327)$ \\
Household-head education code / 9 & $-0.8462$ & $-0.0364$ & $-0.0854$ \\
            & $(0.8970)$ & $(0.0496)$ & $(0.0547)$ \\
Elderly member present & $0.2935$ & $0.0224$ & $0.0203$ \\
            & $(0.2341)$ & $(0.0174)$ & $(0.0190)$ \\
\bottomrule
\end{tabular}

\vspace{0.35em}
\begin{minipage}{0.96\textwidth}
\footnotesize\textit{Note:} All specifications use 3,102 households, including 265 observed entrants, incorporate all seven household characteristics, and use globally normalized survey weights. Parentheses report county-clustered standard errors, covering 107 counties and 125 comparison sets. Logit coefficients describe changes in log odds, whereas LPM coefficients describe linear changes in entry probability. The table shows the three focal characteristics used in subsequent directional constraint tests; Appendix~\ref{app:simple-entry-benchmarks} reports full estimates.
\end{minipage}
\end{table}

The household-head education coefficients are negative and the elderly-member coefficients are positive across all three models, but these estimates are less precise than those for income rank. In the pooled Logit model, the county-clustered p-values are 0.348 and 0.213, respectively. The negative income-rank relationship is therefore clearer across these regression specifications. Section 5.2 examines whether the coefficient directions of these household characteristics remain stable when predicted food and education gains enter the allocation score.

\subsection{Household Relationships after Including Predicted Gains}

The BBK model combines seven household characteristics with predicted food and education gains to explain observed Dibao entry. Household characteristics enter the household component, while predicted gains affect the score through the gain-adjustment term. We examine whether the coefficients on income rank, household-head education, and elderly-member status retain their signs when gain coefficients vary and the household-component parameters are reoptimized.

We focus on 57 solutions in the specified 1\% raw-loss-tolerance region whose gain-adjustment terms are strictly positive for all 3,102 households, $G_i(\beta)>0$. Under this condition, holding other factors fixed, a higher household component gives a higher complete allocation score. The household-component coefficient signs can therefore be used to describe the corresponding allocation directions.

Across all 57 solutions, income-rank and household-head education coefficients are negative, while the elderly-member coefficient is positive: 57 of 57 solutions retain each sign. Thus, holding the other household characteristics fixed, lower income rank, lower household-head education, and the presence of an elderly member each correspond to a higher household component. These directions are unchanged over the evaluated food and education gain-parameter combinations.

{\emergencystretch=1em
A wider set of 603 evaluated parameter coordinates satisfying the specified 5\% raw-loss-tolerance criterion exhibits the same predominant patterns: income-rank coefficients are negative in 603/603 solutions, head education in 599/603, and the elderly-member coefficient is positive in 591/603 (Appendices~\ref{app:joint-geometry} and~\ref{app:strict-positivity}). The 57 strictly positive solutions support the directional interpretation between the household component and complete score; the 603-coordinate results summarize coefficient signs over a broader parameter region. The standard errors in Table~\ref{tab:entry-regressions} indicate statistical precision in the receipt regressions, while these sign frequencies describe stability across evaluated structural parameter solutions.\par}

\subsection{Replacement Cost of the Income Relationship and Its Institutional Implications}

Section 5.2 examined whether household-component coefficients retain their directions as predicted-gain parameters change. We next ask whether other household characteristics and predicted gains can compensate for reversing one of these relationships. We separately constrain the coefficients on income rank and household-head education to be nonnegative and that on elderly-member status to be nonpositive. Each restriction is imposed individually, while the remaining household-component parameters and both gain coefficients are reoptimized. We require the gain-adjustment term to be strictly positive for all 3,102 households.

Replacement cost is measured by the share of fit improvement sacrificed in Equation~\eqref{eq:fit-sacrifice}.

Here $L^*$ is the lowest recorded full-model loss, $L^{\mathrm{eq}}$ the loss of the equal-score benchmark within comparison sets, and $L$ the loss after imposing a directional restriction and reoptimizing. The numerator is the additional loss relative to the full model, while the denominator is the improvement of the full model over the equal-score benchmark. A higher $r(L)$ indicates that the restricted relationship is harder for the other parameters to replace.

\begin{table}[htbp]
\centering
\caption{Household-Component Directions and the Cost of Opposite-Direction Restrictions}
\label{tab:directional-constraints}
\small
{\renewcommand{\arraystretch}{1.12}\setlength{\tabcolsep}{4pt}
\begin{tabular}{@{}>{\raggedright\arraybackslash}p{2.75cm}>{\centering\arraybackslash}p{1.70cm}>{\centering\arraybackslash}p{2.05cm}>{\raggedright\arraybackslash}p{3.00cm}>{\centering\arraybackslash}p{1.95cm}@{}}
\toprule
\scriptsize\shortstack[l]{Household\\characteristic} & \scriptsize\shortstack{Direction in\\57 solutions} & \scriptsize\shortstack{Solutions\\retaining\\direction} & \scriptsize\shortstack[l]{Opposite-direction\\restriction} & \scriptsize\shortstack{Fit improvement\\sacrificed} \\
\midrule
Income rank & Negative & 57/57 & Coefficient nonnegative & 33.6\% \\
Household-head education & Negative & 57/57 & Coefficient nonnegative & 5.4\% \\
Elderly member present & Positive & 57/57 & Coefficient nonpositive & 4.6\% \\
\bottomrule
\end{tabular}
}

\vspace{0.35em}
\begin{minipage}{0.96\textwidth}
\footnotesize\textit{Note:} Direction frequencies are drawn from 57 solutions in the specified 1\% raw-loss-tolerance region with strictly positive gain-adjustment terms for all households. Their raw total loss exceeds the baseline minimum by at most 1\%; expressed under Equation~\eqref{eq:fit-sacrifice}, the largest share of fit improvement sacrificed is 19.0\% ($r=0.190$). These measures have different denominators. Opposite-direction costs are based on separate continuous constrained searches and report the lowest-loss feasible solutions recorded. All comparisons hold the structural sample and both gain predictions fixed and reoptimize all unrestricted parameters.
\end{minipage}
\end{table}

Table~\ref{tab:directional-constraints} shows that requiring the income-rank coefficient to be nonnegative sacrifices about 33.6\% of the original in-sample fit improvement, even after the other six household coefficients and both predicted-gain coefficients are reoptimized. The corresponding costs are 5.4\% for household-head education and 4.6\% for elderly-member status. Although all three directions persist across the 57 strictly positive solutions, the negative income relationship is substantially harder to replace.

Dibao eligibility relies on household per capita income, assets, and local minimum-living standards to identify households facing basic-livelihood hardship. Our analysis uses households' relative ranks in surveyed total household income to study observed receipt. Section 5.1 found that lower income is associated with a higher probability of Dibao entry. This subsection further shows that the negative income-related allocation pattern retains a high replacement cost when other household characteristics and predicted gains are allowed to adjust. This provides additional evidence on the role of economic hardship in observed Dibao allocation.

\section{Predicted Policy Gains, Household Rankings, and Allocation Outcomes}

\subsection{Relative Roles of Predicted Food and Education Gains}

Section 5 established that the income-related allocation relationship is difficult to replace with other factors. We now examine predicted food and education gains: after seven household characteristics are already included in the model, how much additional information do the two predicted gains provide about observed Dibao receipt?

\begin{figure}[htbp]
\centering
\includegraphics[width=\textwidth]{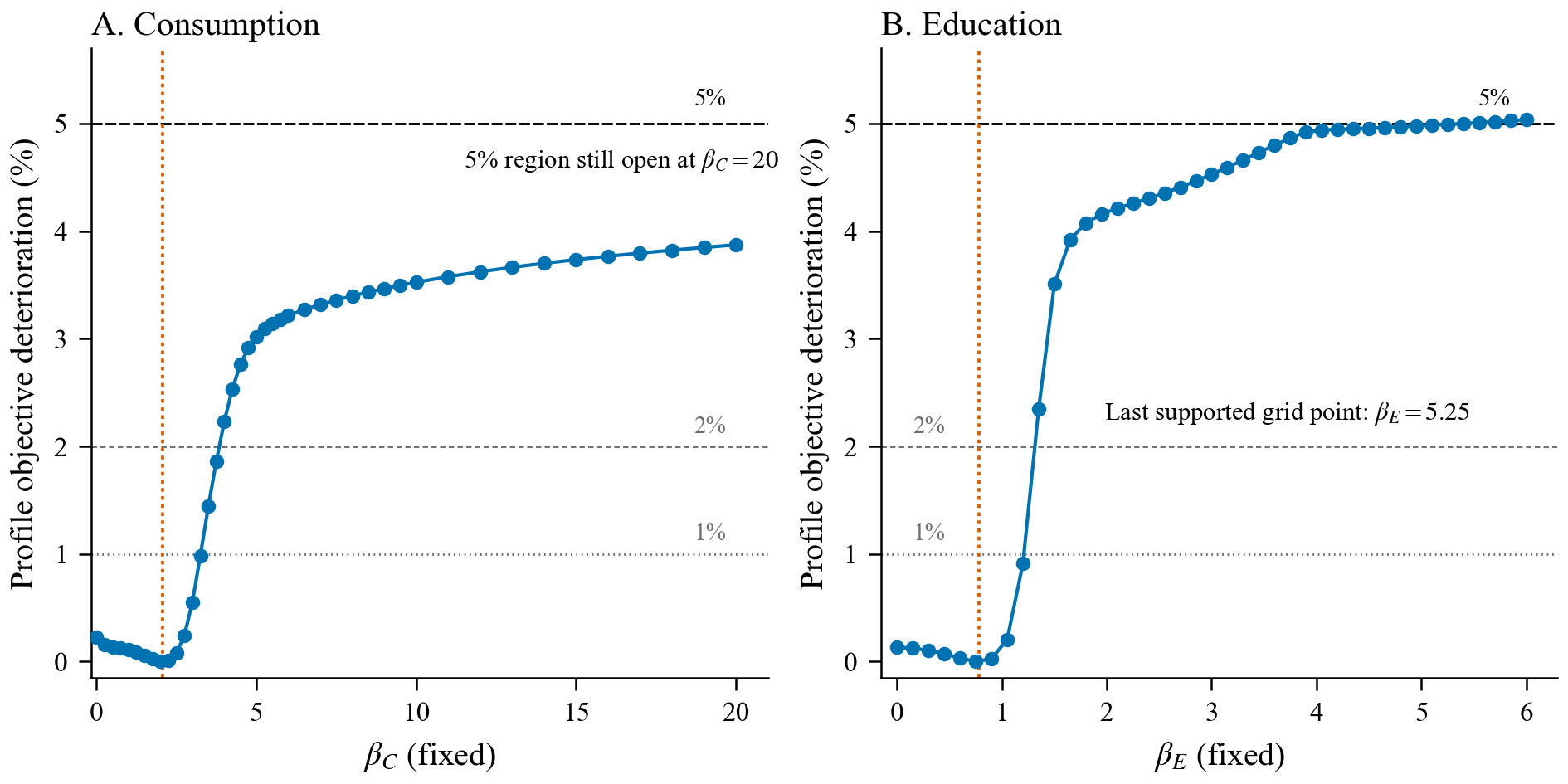}
\caption{Fit Profiles for Food and Education Gain Parameters}
\label{fig:gain-profiles}

\vspace{0.25em}
\begin{minipage}{0.96\textwidth}
\footnotesize\textit{Note:} The left and right panels hold the food and education gain coefficients fixed, respectively, and reoptimize the other gain coefficient and seven household-component parameters. Vertical dashed lines mark the coefficients associated with the lowest recorded baseline loss. The vertical axis shows the percentage increase in profile loss relative to the baseline minimum. Horizontal lines mark specified tolerances of 1\%, 2\%, and 5\% increases in raw total loss. These are not confidence intervals and use a different denominator from Equation~\eqref{eq:fit-sacrifice}. See Appendix~\ref{app:one-dimensional-profiles} for the full search ranges.
\end{minipage}
\end{figure}

In the baseline search, the lowest recorded loss corresponds to a food-gain coefficient of 2.078840 and an education-gain coefficient of 0.782621. Figure~\ref{fig:gain-profiles} holds each coefficient fixed in turn and reoptimizes the other gain coefficient and seven household-component parameters to trace the loss profile. Within the specified 5\% raw-loss-tolerance region, the one-dimensional ranges for both coefficients include zero. Thus, smaller or even zero values of either coefficient can maintain similar fit when the other parameters are reoptimized. Appendix~\ref{app:one-dimensional-profiles} reports the full profile coordinates.

To measure their joint contribution, we set both predicted-gain coefficients to zero and reoptimize all seven household-component coefficients. The recorded loss of this zero-gain model is 1.733 higher than that of the full model. Relative to the full model's improvement over the equal-score-within-set benchmark, this sacrifices 4.36\% of the original fit improvement. The zero-gain model therefore retains approximately 95.6\%. Once the seven household characteristics enter the household component, the two predicted gains make a relatively small additional contribution to in-sample fit.

We also examine the relative ordering of the gain coefficients. The baseline food coefficient exceeds the education coefficient. When the education coefficient is required to exceed the food coefficient by at least 0.05, with gain adjustments strictly positive for every household, the recorded feasible solution sacrifices only about 4.96\% of the original fit improvement. Reversing the coefficients' order can therefore yield fit close to the baseline. On the specified scales for food and education outcomes, observed Dibao receipt provides relatively weak restrictions on the gains' relative coefficients.

This subsection compares how gain coefficients explain observed Dibao receipt. Evaluating the welfare value of improvements in food consumption or education spending additionally depends on households' initial circumstances, how the two outcomes are valued, and the duration of policy effects.

\subsection{Shared Information and Household Rankings}

Section 6.1 shows that the model can maintain similar fit even when the food and education gain coefficients change substantially. An important reason is that the predicted gains and the household component use the same seven household characteristics. Both gain predictions can be expressed as linear functions of those characteristics and a constant, whereas the household component is an exponential function of the seven characteristics (Appendix~\ref{app:gain-dependence}). A household characteristic can therefore affect the complete allocation score through two channels: the household component and the predicted gains. After changing the gain coefficients, reoptimizing the household-component parameters may offset part of the score change.

We assess whether this adjustment preserves household rankings using 57 solutions in the specified 1\% raw-loss-tolerance region with strictly positive gain adjustments for every household. For each solution, we compute Spearman rank correlation with the baseline complete ranking within each comparison set and aggregate these correlations using the number of household pairs in each set as weights. The comparison has two steps. First, we hold the baseline household component fixed and replace only the food and education gain coefficients. Second, we use the household component reoptimized for that gain-coefficient pair. The median rank correlation is 0.853 in the first step and increases to 0.960 in the second. Thus, after reoptimization, the alternative model's household ordering is substantially closer to the baseline ordering. Appendix~\ref{app:evidence-sources} details the calculations.

Different combinations of household-component coefficients and gain coefficients can therefore generate similar household rankings. Based on observed Dibao entry alone, it is difficult to distinguish whether a household's higher allocation score primarily reflects its household characteristics or its predicted food and education gains. The county-held-out diagnostic in Section 4.4 also found that the evaluated heterogeneous-gain predictions did not outperform the constant-effect benchmark on average outside the training counties. That diagnostic was conducted separately from the implemented household-level gain-generation procedure; the structural model consistently uses the original fixed gain predictions.

\subsection{Simulated Recipient Lists and Household-Group Composition}

Does a similar household ranking imply that the same households are selected? Following the fixed-quota rule in Section 3.5, we select the highest-scoring households in each comparison set, with the quota equal to that set's observed number of Dibao entrants. There are 265 simulated places in total. Of the 265 households selected by the baseline model, 77 were observed entrants, giving an overlap of 29.1\%. We therefore use the simulations below to compare recipient lists and their composition across parameter combinations, rather than treating simulated selections as the actual receipt list.

\begin{figure}[htbp]
\centering
\includegraphics[width=\textwidth]{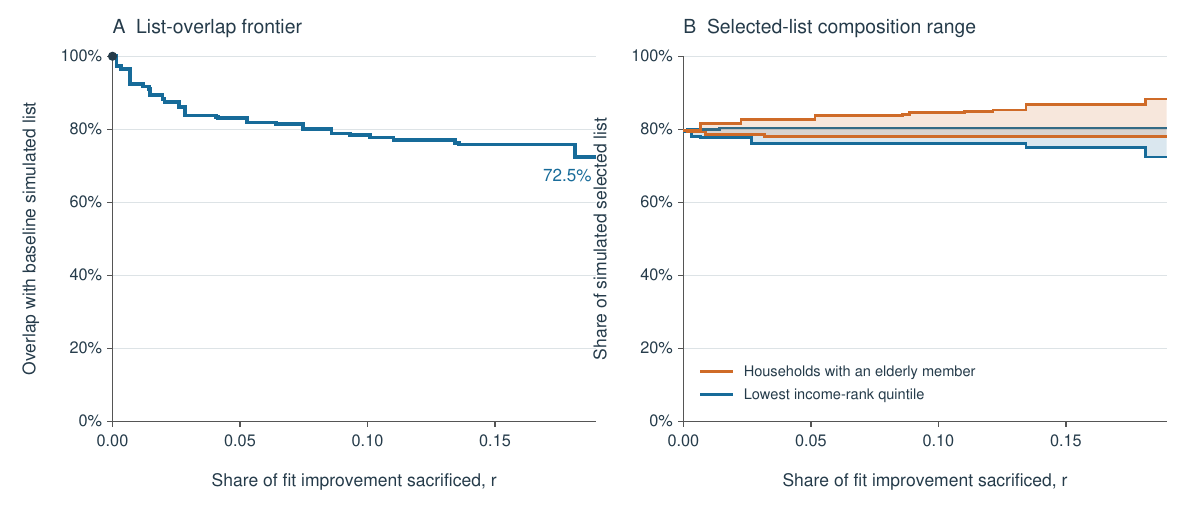}
\caption{List Overlap and Group Composition across Strictly Positive Solutions}
\label{fig:list-frontiers}

\vspace{0.25em}
\begin{minipage}{0.96\textwidth}
\footnotesize\textit{Note:} The figure includes 57 solutions in the specified 1\% raw-loss-tolerance region with strictly positive gain adjustments for every household, plus the baseline solution at $r=0$. The horizontal axis is the share of fit improvement sacrificed, as defined in Equation~\eqref{eq:fit-sacrifice}, with a maximum of 0.190. The left panel shows the minimum overlap with the baseline simulated list; the right panel shows the ranges of list shares for households in the lowest income-rank quintile and households with an elderly member. Overlaps and shares use 265 simulated places as the denominator; vertical axes range from 0\% to 100\%. Shading indicates the range across evaluated solutions, not confidence intervals. Table~\ref{tab:list-composition} and Table L.1 cover only the 57 solutions and do not additionally count the baseline.
\end{minipage}
\end{figure}

Across the 57 strictly positive solutions in the specified 1\% raw-loss-tolerance region, households in the lowest income-rank quintile make up 72.5\%--80.4\% of simulated lists, and households with an elderly member make up 78.1\%--88.3\%. The solutions consistently select high shares of these groups, but the identity of selected households varies. The minimum overlap with the baseline simulated list is 72.5\%, while the minimum overlap between any two alternative lists is 66.8\%. A total of 154 households from the baseline list remain selected in all 57 solutions, or 58.1\% of the 265 places. Figure~\ref{fig:list-frontiers} and Table~\ref{tab:list-composition} summarize the changes in list overlap and group composition.

\begin{table}[htbp]
\centering
\caption{Simulated-List Stability and Group Composition in 57 Strictly Positive Solutions}
\label{tab:list-composition}
\small
\begin{tabular}{@{}p{11.3cm}r@{}}
\toprule
Measure & Result \\
\midrule
Minimum overlap with baseline simulated list & 72.5\% \\
Minimum overlap between any two simulated lists & 66.8\% \\
Share of baseline list retained in all solutions & 58.1\% \\
Always selected / never selected / changing selection status & 154 / 2,698 / 250 \\
Share of households in lowest income-rank quintile & 72.5\%--80.4\% \\
Share of households with an elderly member & 78.1\%--88.3\% \\
Selection-change events within one rank of baseline cutoff & 86.3\% \\
\bottomrule
\end{tabular}

\vspace{0.35em}
\begin{minipage}{0.96\textwidth}
\footnotesize\textit{Note:} All 57 solutions have strictly positive gain adjustments for the 3,102 households and raw total losses within 1\% of the baseline minimum; under Equation~\eqref{eq:fit-sacrifice}, the maximum sacrificed fit improvement is 19.0\% ($r=0.190$). The baseline solution is not additionally included in the table. List-overlap measures, the common-retention share, and group shares use 265 places as the denominator; selection-status categories refer to all 3,102 households. The lowest quintile is formed using the income ranking in the structural sample. Distances from the selection cutoff are calculated for household-by-parameter selection-change events, so a household may be counted repeatedly. See Appendix~\ref{app:list-frontier} for definitions and full results.
\end{minipage}
\end{table}

Changes in individual selection are concentrated near the quota cutoffs. Across the 57 solutions, there are 4,712 household-by-parameter selection-change events relative to the baseline list; 86.3\% concern households within one rank of the baseline cutoff, while 6.1\% are at least three ranks away. Each household's selection status in each parameter solution is counted separately. With quotas fixed within comparison sets, small changes in ordering near a cutoff can change which individual household is selected even when the overall shares of low-income households and households with elderly members remain similar. Stable group composition and changes in individual recipient identities can therefore occur together.

\subsection{Fit Criteria, Geographic Weights, and Allocation Sensitivity}

This section varies fit criteria and geographic weights to assess the stability of household-characteristic coefficient signs and simulated recipient lists. We first compare five full-sample specifications: the baseline model and four alternatives that change comparison-set weights, use a ranking loss, or adjust the softmax temperature. All five yield negative coefficients on income rank and household-head education and a positive coefficient on elderly-member status. The four alternative specifications overlap with the baseline simulated list by 74.0\%--90.6\%; overlap with observed entrants across all five specifications ranges from 27.9\% to 32.1\%. We check the gain-adjustment terms after estimation. Three households have nonpositive gain adjustments in the temperature-2 specification, while the other four specifications satisfy positivity for all households. Appendix~\ref{app:loss-sensitivity} reports the complete results.

We next reweight counties to assess how the contribution of different geographic areas affects the results. The five specifications use the same 30 county-level weight perturbations, yielding 150 numerically valid estimates. Within each specification, the income-rank coefficient is negative in 29--30 of the 30 reweightings; the median overlap with the fixed baseline simulated list ranges from 62.6\% to 72.8\%. One perturbation of the equal-weight-across-sets softmax specification reverses the sign of the income coefficient, producing a simulated list that shares only 11 households (4.2\%) with the baseline list. That solution also has 144 households with nonpositive gain adjustments. For those households, the direction of the household-component coefficients cannot be used to interpret the direction of the complete score. Appendix~\ref{app:loss-sensitivity} provides the other coefficients, rankings, and numerical diagnostics.

Finally, list stability depends on the permitted fit loss and parameter range. Under the baseline fit criterion, among evaluated solutions in which both gain coefficients are in [0,10] and at least 95\% of the baseline improvement in in-sample fit is retained ($r\leq 0.05$), overlap with the baseline simulated list is never below 83.0\%. In contrast, the broad region allowing raw total loss to increase by no more than 5\% over its baseline minimum can sacrifice approximately 96.1\% of the original fit improvement. Its 603 parameter vectors also include expanded coordinates outside the main parameter range and can generate larger list differences. These comparisons use different loss denominators and parameter spaces; see Appendices~\ref{app:fixed-quota-sensitivity} and~\ref{app:list-frontier} for details.

Overall, the negative income-rank relationship remains stable under most alternative specifications and county reweightings, whereas individual simulated lists are more sensitive to gain coefficients, fit requirements, and geographic weights. This is consistent with Section 6.3: the aggregate composition of recipient groups can be similar even as the selection status of individual households changes.

\section{Discussion}

\subsection{Household Economic Conditions in Observed Dibao Allocation}

Standard Logit and linear probability model (LPM) regressions show that, after controlling for other household characteristics, lower-income households were more likely to receive Dibao in 2019. The BBK model asks a further question: when other household characteristics and predicted gains in food consumption and education spending are considered together, can other factors substitute for the negative relationship between income and Dibao receipt? Requiring the gain-adjustment term to be positive for every household, constraining the income-rank coefficient to be nonnegative and reoptimizing the remaining parameters sacrifices approximately 33.6\% of the original improvement in in-sample fit. The corresponding losses from imposing opposite-sign restrictions on household-head education and elderly-member status are 5.4\% and 4.6\%. Among these three relationships, the negative income relationship is the most difficult to replace through adjustments to other parameters.

This finding adds to our understanding of how Dibao assistance is allocated in practice. Earlier studies have focused on program coverage, recipients’ poverty status, and targeting performance (Chen et al., 2006; Golan et al., 2017). We additionally examine how far other household characteristics or predicted policy gains can substitute for the income–receipt relationship. The negative income relationship remains stable even when other model parameters are allowed to adjust. Since Dibao uses household economic circumstances as a central eligibility criterion, this result is consistent with its role in protecting the basic livelihoods of households facing hardship. It also raises a broader policy question: how does prioritizing households facing the greatest current hardship differ from prioritizing households more likely to achieve sustained poverty exit with assistance?

\subsection{The Policy Objectives of Basic-Livelihood Protection and Sustained Poverty Exit}

Social assistance can be allocated according to two different principles. Need-based targeting prioritizes households according to their current economic hardship, with the aim of protecting the basic livelihoods of those most in need. Poverty-exit-oriented targeting instead gives priority to households whose income and earning capacity are more likely to improve with assistance, allowing them to rely less on transfers over time. The two principles may select different recipients. Households with the lowest incomes often have the most urgent need for support, yet may not be the most likely to escape poverty in the short run. Some households facing less severe hardship may be better positioned to achieve sustained poverty exit with assistance. Evaluating the allocation of social assistance therefore begins with identifying the policy objective being prioritized.

One possible reason why development-oriented assistance may produce different sustained-poverty-exit effects across households is the interaction between credit constraints and thresholds for productive investment. For a disadvantaged household already close to such a threshold, a transfer may finance an otherwise unattainable investment and support continued income growth. A household in extreme poverty may instead need to use the same transfer first for basic consumption, leaving it unable to cross the investment threshold in the short run (Galor and Zeira, 1993; Balboni et al., 2022). This mechanism is a theoretical possibility rather than a causal channel tested in this study.

The distinction also matters for social welfare evaluation. Haushofer et al. (2025) compare allocations based on deprivation and predicted policy impacts in a Kenyan cash-transfer program. Their analysis highlights the difference between the size of an improvement and its social welfare value. For very low-income households, the same cash transfer may have greater marginal welfare value. For households better positioned to improve their economic circumstances, assistance may lead to larger income gains or other policy effects. Which allocation achieves higher social welfare depends on households' initial circumstances, the actual effects of assistance, and the weights assigned to different households and outcomes in the social welfare objective.

Our results provide empirical evidence for understanding Dibao's need-based orientation. Even after other model parameters are reoptimized, the negative relationship between household income and Dibao receipt remains difficult to replace. This is consistent with the program's role in protecting the basic livelihoods of households facing economic hardship. It also raises a further question: even if prioritizing the poorest households does not produce the largest number of households exiting poverty in the short run, why might this allocation still maximize social welfare? The next section discusses the marginal welfare value of assistance to the poorest households and the potential long-run benefits of basic-livelihood protection.

\subsection{The Multidimensional and Long-Run Welfare Value of Basic-Livelihood Protection}

Prioritizing the poorest households may yield substantial social welfare gains because a household's initial economic circumstances affect the marginal utility of assistance. For families with very low incomes and difficulty meeting basic needs, the same transfer amount may deliver a larger improvement in well-being. If the social welfare criterion places greater weight on these improvements, allocating scarce assistance to the poorest can yield higher aggregate welfare even when helping other households would generate more short-run exits from poverty. Comparing allocation rules therefore requires accounting for households' initial living standards, the actual effects of assistance, and the weights assigned to different households' welfare.

Basic-livelihood protection may also generate longer-run benefits by reducing economic risk. For households with unstable incomes and limited prospects of achieving economic independence in the short run, regular Dibao benefits can sustain essential consumption, ease the strain caused by income interruptions, and reduce concern over meeting future basic needs. Stable support may also help households plan spending on education, employment, and other long-term needs. Gao (2017) notes that the coverage and level of Dibao benefits influence its poverty-reduction performance, highlighting the importance of sufficient and sustained assistance.

Improved economic security may create more opportunities for disadvantaged households to participate in social life. When basic needs can be met more reliably, households may have more time and resources for community activities, public affairs, and social interaction. Using data on low-income households in China, Hu et al. (2024) find that Dibao promotes participation in participatory governance and the coproduction of public services, although the improvement in interpersonal social participation is more limited. Gao (2017), by contrast, reports associations between Dibao receipt and less social and leisure activity and lower subjective well-being. These studies show that assistance can affect outcomes beyond economic conditions, with effects that may differ across forms of social participation and lived experience.

The administration of Dibao may also affect households' trust in public institutions. Fair eligibility reviews, regular benefit payments, and reliable implementation can strengthen expectations of public support. Reviewing evidence on social transfers, Kosec and Mo (2025) document both positive and negative effects on institutional trust, governance, and social cohesion, depending on program design and implementation. Pan (2020) shows in the Chinese Dibao context that households excluded from assistance may become dissatisfied when allocations also serve other governance objectives, affecting how they evaluate the government. The longer-run welfare value of assistance therefore also depends on perceived fairness in allocation and reliability in implementation.

Taken together, sustained poverty-exit effects and social welfare value are distinct evaluation dimensions. Even when the poorest households are less likely to achieve sustained poverty exit in the short run, protecting their basic livelihoods may have greater social welfare value: the same transfer may yield a larger marginal welfare gain, while stable support may reduce economic risk, support social participation, and affect institutional trust. These possibilities depend on actual program effects and social welfare weights and are not causal mechanisms identified by our analysis. They nevertheless help explain why the need-based orientation indicated by the difficult-to-replace negative income relationship may have substantial welfare value. The next section discusses its implications for China's Dibao system and common-prosperity policies.

\subsection{Dibao's Institutional Role and Common Prosperity}

In China's social-assistance system, Dibao plays a central role in safeguarding basic living standards for households facing hardship. Household income, assets, and other economic conditions are key eligibility criteria, and households meeting local standards may apply for assistance under prescribed procedures. Employment services, skills training, and productive assistance, meanwhile, primarily help eligible disadvantaged households strengthen their ability to work and earn income. Dibao provides basic-livelihood protection, while development-oriented assistance supports further improvements in household economic circumstances. The two types of support serve different functions and can be linked according to household needs.

This institutional arrangement is consistent with China's policy agenda for common prosperity. The 2020 Opinions on Reforming and Improving the Social Assistance System emphasize protecting the basic rights and interests of disadvantaged people and enabling them to share in the gains from development (General Office of the CPC Central Committee and General Office of the State Council, 2020). The 2021 Opinions on Supporting Zhejiang's High-Quality Development as a Demonstration Zone for Common Prosperity incorporate both stronger development capacity among low-income groups and improvements to the social-assistance system (CPC Central Committee and State Council, 2021). These two documents, issued after the study's 2019 observation of Dibao receipt, reflect an approach that combines basic-livelihood protection with better development opportunities. Reliable social assistance provides essential support, while employment services and training create opportunities for households able to strengthen their earning capacity.

This study finds that the negative relationship between income and observed Dibao receipt is difficult to replace with other factors, consistent with the program's role in safeguarding basic livelihoods among households in financial hardship. Fixed-quota simulations further show that different parameter combinations can select similar proportions of low-income households while selecting different individual households. Understanding actual Dibao allocation therefore requires attention both to receipt across groups with different economic circumstances and to how individual households obtain assistance through application, verification, and approval. Combining administrative review records, actual transfer amounts, and subsequent household economic outcomes could help explain local recipient selection and how basic-livelihood protection connects with development-oriented assistance as household conditions improve.

\section{Conclusion}

Using the BBK revealed-allocation framework, this paper examines how seven household characteristics and predicted gains in food consumption and education spending explain observed Dibao receipt in China. The negative relationship between income and receipt is difficult to offset by changing other model parameters. When the gain-adjustment term is positive for every household, constraining the income-rank coefficient to be nonnegative substantially worsens fit with observed receipt, even after other parameters are reoptimized. By contrast, removing both predicted gains has a relatively small effect on fit. These findings identify income as a particularly important feature of observed Dibao allocation and illustrate the value of comparing whether different model factors can substitute for one another.

The findings provide evidence relevant to Dibao's need-based orientation. The persistent negative income relationship is consistent with the program's role in protecting the basic livelihoods of economically disadvantaged households. It also highlights the importance of the objective used to evaluate social-assistance allocation. Prioritizing households facing the greatest current hardship and seeking stronger near-term poverty-exit outcomes can imply different recipient selections and call for different welfare assessments.

From a welfare perspective, prioritizing the poorest households can have substantial value. An equal transfer may yield a larger marginal welfare gain for households with fewer initial resources, while reliable basic-livelihood protection may also generate lasting economic and social benefits. Under particular welfare weights and policy effects, need-based targeting can therefore be consistent with maximizing social welfare. By connecting evidence on observed Dibao allocation with this broader welfare question, the study offers a perspective on how social-assistance objectives should be evaluated. Future work combining administrative eligibility records, actual transfer amounts, and long-run household panel data could shed further light on recipient selection and longer-run welfare outcomes.

\section*{References}

\refentry{Alatas, V., Banerjee, A., Hanna, R., Olken, B.A., Tobias, J., 2012. Targeting the poor: Evidence from a field experiment in Indonesia. \textit{American Economic Review} 102(4), 1206--1240. \url{https://doi.org/10.1257/aer.102.4.1206}}

\refentry{Balboni, C., Bandiera, O., Burgess, R., Ghatak, M., Heil, A., 2022. Why do people stay poor? \textit{The Quarterly Journal of Economics} 137(2), 785--844. \url{https://doi.org/10.1093/qje/qjab045}}

\refentry{Björkegren, D., Blumenstock, J.E., Knight, S., 2026. What do policies value? \textit{The Review of Economic Studies} 93(4), 2424--2450. \url{https://doi.org/10.1093/restud/rdaf089}}

\refentry{Bourguignon, F., Spadaro, A., 2012. Tax--benefit revealed social preferences. \textit{The Journal of Economic Inequality} 10(1), 75--108. \url{https://doi.org/10.1007/s10888-010-9153-0}}

\refentry{Brown, C., Ravallion, M., van de Walle, D., 2018. A poor means test? Econometric targeting in Africa. \textit{Journal of Development Economics} 134, 109--124. \url{https://doi.org/10.1016/j.jdeveco.2018.05.004}}

\refentry{Chen, S., Ravallion, M., Wang, Y., 2006. Di Bao: A guaranteed minimum income in China's cities? World Bank Policy Research Working Paper No. 3805, January. World Bank. \url{https://documents.worldbank.org/en/publication/documents-reports/documentdetail/399721468019146622}}

\refentry{Chernozhukov, V., Chetverikov, D., Demirer, M., Duflo, E., Hansen, C., Newey, W., Robins, J., 2018. Double/debiased machine learning for treatment and structural parameters. \textit{The Econometrics Journal} 21(1), C1--C68. \url{https://doi.org/10.1111/ectj.12097}}

\refentry{CPC Central Committee and State Council, 2021. Opinions on supporting Zhejiang in high-quality development and building a demonstration zone for common prosperity. Document dated May 20; published June 10. In Chinese. {\def\UrlBreaks{\do\/\do\.\do\0\do\1\do\2\do\3\do\4\do\5\do\6\do\7\do\8\do\9}\url{https://www.gov.cn/gbgl/92e7fad3d97742f1b1b2beb1fda9c049/files/8525ced0638e4480a1368790d9433924.pdf}}}

\refentry{Galor, O., Zeira, J., 1993. Income distribution and macroeconomics. \textit{The Review of Economic Studies} 60(1), 35--52. \url{https://doi.org/10.2307/2297811}}

\refentry{Gan, L., Yin, Z., Jia, N., Xu, S., Ma, S., Zheng, L., 2014. \textit{Data You Need to Know About China: Research Report of China Household Finance Survey 2012}. Springer, Berlin/Heidelberg. \url{https://doi.org/10.1007/978-3-642-38151-5}}

\refentry{Gao, Q., 2017. \textit{Welfare, Work, and Poverty: Social Assistance in China}. Oxford University Press. \url{https://doi.org/10.1093/oso/9780190218133.001.0001}}

\refentry{General Office of the CPC Central Committee and General Office of the State Council, 2020. Opinions on reforming and improving the social assistance system. Published August 25. In Chinese. \url{https://www.beijing.gov.cn/zhengce/zhengcefagui/202008/t20200826_1990925.html}}

\refentry{Golan, J., Sicular, T., Umapathi, N., 2017. Unconditional cash transfers in China: Who benefits from the rural minimum living standard guarantee (Dibao) program? \textit{World Development} 93, 316--336. \url{https://doi.org/10.1016/j.worlddev.2016.12.011}}

\refentry{Han, H., Gao, Q., 2019. Community-based welfare targeting and political elite capture: Evidence from rural China. \textit{World Development} 115, 145--159. \url{https://doi.org/10.1016/j.worlddev.2018.11.010}}

\refentry{Hanna, R., Olken, B.A., 2018. Universal basic incomes versus targeted transfers: Anti-poverty programs in developing countries. \textit{Journal of Economic Perspectives} 32(4), 201--226. \url{https://doi.org/10.1257/jep.32.4.201}}

\refentry{Haushofer, J., Niehaus, P., Paramo, C., Miguel, E., Walker, M., 2025. Targeting impact versus deprivation. \textit{American Economic Review} 115(6), 1936--1974. \url{https://doi.org/10.1257/aer.20221650}}

\refentry{Hendren, N., 2020. Measuring economic efficiency using inverse-optimum weights. \textit{Journal of Public Economics} 187, 104198. \url{https://doi.org/10.1016/j.jpubeco.2020.104198}}

\refentry{Hu, H., He, H., Li, Y., Zhang, K., Hu, X., 2024. Does cash transfer program improve recipients' social participation? Evidence from low-income households data in China. \textit{International Journal of Social Welfare} 33(4), 1057--1079. \url{https://doi.org/10.1111/ijsw.12670}}

\refentry{Kosec, K., Mo, C.H., 2025. The unintended consequences of social transfer programs: Effects on conflict, governance, and social cohesion. \textit{Annual Review of Resource Economics} 17, 467--491. \url{https://doi.org/10.1146/annurev-resource-112923-094210}}

\refentry{Kuhn, L., Brosig, S., Zhang, L., 2016. The brink of poverty: Implementation of a social assistance programme in rural China. \textit{Journal of Current Chinese Affairs} 45(1), 75--108. \url{https://doi.org/10.1177/186810261604500105}}

\refentry{Li, M., Walker, R., 2021. Need, justice and central--local relations: The case of social assistance in China. \textit{Public Administration} 99(1), 87--102. \url{https://doi.org/10.1111/padm.12689}}

\refentry{Ministry of Civil Affairs, 2012. Notice issuing the measures for review and approval of the minimum living standard guarantee (trial). Minfa [2012] No. 220, December 12. In Chinese.}

\refentry{Nie, X., Wager, S., 2021. Quasi-oracle estimation of heterogeneous treatment effects. \textit{Biometrika} 108(2), 299--319. \url{https://doi.org/10.1093/biomet/asaa076}}

\refentry{Pan, J., 2020. \textit{Welfare for Autocrats: How Social Assistance in China Cares for Its Rulers}. Oxford University Press. \url{https://doi.org/10.1093/oso/9780190087425.001.0001}}

\refentry{Ravallion, M., 2008. Miss-targeted or miss-measured? \textit{Economics Letters} 100(1), 9--12. \url{https://doi.org/10.1016/j.econlet.2007.10.020}}

\refentry{Ravallion, M., 2009. Decentralizing eligibility for a federal antipoverty program: A case study for China. \textit{The World Bank Economic Review} 23(1), 1--30. \url{https://doi.org/10.1093/wber/lhn023}}

\refentry{State Council, 2014. \textit{Interim Measures for Social Assistance}. State Council Decree No. 649, February 21; effective May 1; amended 2019. In Chinese.}

\refentry{Tamer, E., 2010. Partial identification in econometrics. \textit{Annual Review of Economics} 2, 167--195. \url{https://doi.org/10.1146/annurev.economics.050708.143401}}

\clearpage
\begin{center}
{\Large\bfseries Online Appendix}\\[0.45em]
{\large Who Is Prioritized? Current Need versus Predicted Gains\\
in China's Dibao Allocation: Evidence from a Revealed-Allocation Model}
\end{center}

\appendix
\setcounter{table}{0}
\renewcommand{\thetable}{A.\arabic{table}}
\setcounter{figure}{0}
\renewcommand{\thefigure}{A.\arabic{figure}}

\section{Sample Construction, Variables, and First-Stage Diagnostics}
\label{app:sample-construction}

\subsection{Parallel Sample Branches}

The sample begins with the archived W2 analysis frame rather than the original CHFS population. After applying the verified panel-linkage, baseline nonreceipt, and fixed-common-support restrictions in sequence, the prediction population contains 6,132 households. All three analysis branches start from this population. Table~\ref{tab:app-sample-construction} reports each restriction and the corresponding number of entrants.

The structural branch defines comparison sets by county (district) and urban or rural status, retaining sets with at least three households and with both entrants and nonentrants. The screening conditions are
\begin{equation}
\begin{gathered}
m=\text{county/district}\times\text{urban/rural},\\
N_m\geq 3,\qquad 1\leq K_m\leq N_m-1,
\end{gathered}
\tag{A.1}
\end{equation}
where $N_m$ is the number of sampled households in comparison set $m$, and $K_m$ is the number of observed entrants defined from survey-reported receipt status. The final sample contains 125 comparison sets spanning 107 counties (districts). These restrictions ensure that every retained set includes both entrants and nonentrants, allowing the model to fit observed entry from relative differences in household allocation scores within the set. Scores are normalized within each comparison set, and set-level losses are aggregated for estimation, so score levels need not be compared directly across counties. The household-component and predicted-gain coefficients are nevertheless estimated jointly across all comparison sets. The model therefore uses a common parameter vector to describe conditional allocation relationships across locations; it does not assume identical local screening practices. Within-set comparisons determine the geographic scope of the identifying variation, while sensitivity to local implementation is examined separately through geographic reweighting.

\begin{table}[htbp]
\centering
\caption{Sample construction}
\label{tab:app-sample-construction}
\small
\begin{tabular}{lrrrr}
\toprule
Stage & Starting & Removed & Retained & Entrants \\
 & households & households & households & \\
\midrule
Archived W2 analysis frame & --- & --- & 6,651 & 267 \\
Verified panel linkage & 6,651 & 42 & 6,609 & 266 \\
Baseline nonrecipient households & 6,609 & 420 & 6,189 & 266 \\
Fixed common support & 6,189 & 57 & 6,132 & 266 \\
Complete covariates and positive weight & 6,132 & 0 & 6,132 & 266 \\
Food outcome observed & 6,132 & 408 & 5,724 & 252 \\
Education outcome observed & 6,132 & 364 & 5,768 & 248 \\
Structural branch: remove small sets & 6,132 & 58 & 6,074 & 265 \\
Structural branch: remove sets without entrants & 6,074 & 2,972 & 3,102 & 265 \\
\bottomrule
\end{tabular}
\parbox{0.96\textwidth}{\footnotesize\textit{Note:} The food, education, and structural branches each begin separately from the prediction population. Forty small comparison sets contain 58 households; another 212 sets without entrants contain 2,972 households. After the size restriction is imposed, no comparison set consists entirely of entrants. The final 125 comparison sets span 107 counties.}
\end{table}

In the structural sample, the food outcome is observed for 2,934 households, including 251 entrants, and missing for 168 households, including 14 entrants. The education outcome is observed for 2,940 households, including 247 entrants, and missing for 162 households, including 18 entrants. Missingness in the two outcomes may overlap, so the two missing-household counts cannot be added to obtain a deduplicated total. Households with complete covariates but missing realized outcomes can still receive gain predictions, although they do not enter estimation of the corresponding heterogeneous-treatment-effect coefficients.

\subsection{Variable Definitions}

The seven household characteristics enter both the heterogeneous-treatment-effect function in the primary gain-generation procedure and the household component. Table~\ref{tab:app-covariates} reports the scales used in estimation. Income rank follows a fixed survey-weighted empirical distribution, with midranks assigned to ties. The archived materials do not fully recover the earlier reference population used to construct this distribution, so the measure is not described as a within-county percentile or as administrative income. The original education variable is coded from 1 to 9; the archive does not contain the category labels, and none are inferred from the coding range.

\begin{table}[htbp]
\centering
\caption{Household covariates}
\label{tab:app-covariates}
\small
\begin{tabular}{p{0.19\textwidth}p{0.36\textwidth}p{0.36\textwidth}}
\toprule
Variable & Construction & Interpretation \\
\midrule
Income rank & Survey-weighted ranking of household total income & Not administratively assessed per capita income \\
Household size & Number of household members divided by 10 & Coefficient uses the scaled unit \\
Child present & One if a member aged 0--15 is present & Presence indicator, not the number of children \\
Elderly member & One if a member aged 60 or older is present & Does not by itself establish formal priority status \\
Female household head & One if the original sex code of the household head equals 2 & Characteristic of the household head \\
Age of household head & Age divided by 100 & Not age in unscaled years \\
Education of household head & Original code (1--9) divided by 9 & Not years of schooling \\
\bottomrule
\end{tabular}
\end{table}

The food outcome is the 2017--2021 change in the log of real per capita food consumption. The education outcome is the change in the log of one plus real total household spending on education and training. The accompanying data dictionary records each original or inherited field, its use, unit, coding, and any known provenance gap.

The original field-level construction of food consumption is retained below. Food expenditure is divided by household size and deflated by the consumer price index; Equation~\eqref{eq:4-1} in the main text uses $C$, $n$, and $d$ as equivalent shorthand:
\begin{equation}
\begin{gathered}
C_{it}^{\mathrm{real,pc}}
=\frac{\mathrm{food\_total\_annual}_{it}}
{\mathrm{household\_size}_{it}d_t},
\qquad
v_{C_i,t}=\ln\!\left(C_{it}^{\mathrm{real,pc}}\right),\\
\Delta v_{C_i}=v_{C_i,2021}-v_{C_i,2017}.
\end{gathered}
\tag{A.2}
\end{equation}
The original field for the education outcome is \texttt{g1016}, total household spending on education and training. After price deflation, the measure is transformed using $\log(1+x)$ and differenced between 2017 and 2021. Households with zero spending are retained, and the measure is defined at the household-total level.

\subsection{Geographic Identifiers}

Both cluster fields in the structural sample map one-to-one to counties. The grouping variable used for household-level cross-fitting has no township-level administrative interpretation (Table~\ref{tab:app-geo-fields}).

\begin{table}[htbp]
\centering
\caption{Geographic fields in the structural sample}
\label{tab:app-geo-fields}
\small
\begin{tabular}{lrrl}
\toprule
Field & Missing & Unique values & Meaning \\
\midrule
\texttt{x\_cluster\_id} & 0 & 107 & One-to-one mapping to county \\
\texttt{cluster\_id} & 0 & 107 & One-to-one mapping to county \\
\texttt{fold\_group\_id} & 0 & 3,102 & Stable household identifier \\
\bottomrule
\end{tabular}
\end{table}

The available identifiers do not support calculation of the number of townships, the number of households successfully matched to townships, or township-level sample attrition. These quantities are unavailable rather than zero. A township-level comparison would require an anonymized township or subdistrict identifier with a clearly defined baseline that can be linked through household keys. It would then need to distinguish losses due to linkage failure, small comparison sets, and the absence of within-set variation in entry, and to compare geographic definitions on a common retained sample. County-level multiplier reweighting changes each county's contribution to the objective but does not substitute for this exercise.

\subsection{Implemented First-Stage Gain Generation and Independent Out-of-County Diagnostics}

The implemented gain-generation procedure uses five internal folds grouped at the household level to cross-fit the nuisance models. It then fits a linear R-learner on the sample with the corresponding observed outcome and generates $\widehat{\tau}_i$ for the prediction population. The baseline structural model reads these fixed vectors. Out-of-fold nuisance residuals do not imply that the final heterogeneous-treatment-effect coefficients were estimated separately while leaving out each county in turn.

Urban or rural status enters the propensity-score model, the outcome nuisance models, and the county (district)-by-urban/rural comparison sets, but not the linear heterogeneous-treatment-effect function as a separate regressor. That function contains an intercept and the seven household characteristics.

The independent diagnostic introduces five outer county-level folds on the prespecified common support. Test counties do not enter estimation of the propensity-score model, outcome nuisance model, or heterogeneous-treatment-effect coefficients for their fold. Within the training counties, nuisance functions are cross-fitted using household-grouped folds; predictions for test counties use models fitted on the training counties. The propensity score is estimated by $L_2$-penalized logistic regression with $C=0.1$, while the outcome model uses ridge regression with penalty parameter $\alpha=100$. The implemented education-gain procedure considers both ridge regression and histogram gradient boosting when selecting the nuisance model. The diagnostic therefore evaluates the out-of-county performance of a related procedure rather than exactly reproducing the implemented input vectors under independent holdout. It is conditional on the prespecified common-support sample.

For test fold $f$, define $\widetilde y_i=Y_i-\widehat m(X_i)$ and $\widetilde d_i=D_i-\widehat e(X_i)$. The heterogeneous R-loss is
\begin{equation}
R_f^{\mathrm{het}}
=\frac{\sum_{i\in f}w_i\left(\widetilde y_i-\widetilde d_i\widehat\tau_i\right)^2}
{\sum_{i\in f}w_i}.
\tag{A.3}
\end{equation}
Here $w_i$ denotes the survey weight, not the latent household component in the structural model. The constant-effect benchmark is fitted on the training-fold residuals and evaluated on the same test fold. Improvement is defined as $R_f^{\mathrm{const}}-R_f^{\mathrm{het}}$. Calibration uses the no-intercept projection
\begin{equation}
b_f
=\frac{\sum_{i\in f}w_i\left(\widetilde d_i\widehat\tau_i\right)\widetilde y_i}
{\sum_{i\in f}w_i\left(\widetilde d_i\widehat\tau_i\right)^2}.
\tag{A.4}
\end{equation}
This projection neither includes a separate $\widetilde d_i$ term nor centers $\widehat\tau_i$, so it does not isolate the heterogeneity slope from a standard two-term calibration regression. The reported slope is weighted by the number of households in each test fold, $\bar b=\sum_f n_fb_f/\sum_f n_f$. R-loss is aggregated in the same way, with survey weights used within each fold. This aggregation is not a single regression pooling all holdout predictions. Table~\ref{tab:app-county-diagnostics} reports all ten outcome-by-fold records.

\begin{table}[htbp]
\centering
\caption{Fold-level out-of-county diagnostics}
\label{tab:app-county-diagnostics}
\footnotesize
\resizebox{\textwidth}{!}{%
\begin{tabular}{lrrrrrr}
\toprule
Outcome & Fold & Test households & Heterogeneous R-loss & Constant-effect R-loss & Improvement & Calibration \\
\midrule
Food & 0 & 1,117 & 0.964063 & 0.955791 & -0.008271 & -0.569054 \\
Education & 0 & 1,117 & 17.631565 & 17.635450 & 0.003885 & 0.537957 \\
Food & 1 & 1,139 & 0.699752 & 0.696787 & -0.002965 & 0.048886 \\
Education & 1 & 1,156 & 20.976096 & 20.786331 & -0.189764 & -1.682778 \\
Food & 2 & 1,334 & 0.652156 & 0.647853 & -0.004303 & -0.538606 \\
Education & 2 & 1,342 & 16.077100 & 15.964699 & -0.112401 & -1.356591 \\
Food & 3 & 1,099 & 0.808884 & 0.807216 & -0.001667 & 0.135217 \\
Education & 3 & 1,100 & 18.038741 & 18.048952 & 0.010211 & 0.471567 \\
Food & 4 & 1,035 & 0.773887 & 0.776530 & 0.002643 & 0.643531 \\
Education & 4 & 1,053 & 17.991051 & 17.993241 & 0.002189 & 0.529207 \\
\bottomrule
\end{tabular}%
}
\parbox{0.96\textwidth}{\footnotesize\textit{Note:} Improvement is positive in one of five folds for food and in three of five folds for education, but the household-count-weighted average improvement is negative for both outcomes. The table does not report significance tests for the calibration slopes.}
\end{table}

\setcounter{table}{0}
\renewcommand{\thetable}{B.\arabic{table}}
\setcounter{figure}{0}
\renewcommand{\thefigure}{B.\arabic{figure}}

\section{Objective Function, Score Scale, and Numerical Implementation}
\label{app:objective-scale}

\subsection{Objective Function and Score Scale}

The baseline criterion and its derivative with respect to the score can be written as
\begin{equation}
L_m
=W_m\log\!\left(\sum_{j\in m}e^{Q_j}\right)
-\sum_{i\in m}s_iD_iQ_i,
\qquad
\frac{\partial L_m}{\partial Q_j}=W_mp_j-s_jD_j.
\tag{B.1}
\end{equation}
If survey weights do not vary, $W_m=K_m$; otherwise, $W_m$ is the total entrant weight. Multiple entrants share the same denominator and compete for fitted share. The criterion cannot be interpreted mechanically as $K_m(N_m-K_m)$ equally weighted pairwise comparisons.

Under the softmax criterion, the numerical scale of household allocation scores also affects model fit. Adding a common constant to every score in a comparison set leaves normalized shares and the fitting loss unchanged. At a fixed temperature, however, multiplying all scores by the same positive constant generally changes both shares and loss even when the household ranking is preserved. A strictly increasing nonlinear transformation may likewise preserve the ranking while changing fit. The parameters therefore determine both which households receive higher scores and how score gaps enter the estimation objective. The baseline term of one in Section 3.1 normalizes relative valuations within policy value, whereas the softmax temperature governs how score gaps affect fit. Keeping these roles distinct allows us to compare whether a parameter change alters household rankings and why different score representations of the same ranking may fit differently.

For example, consider a comparison set with one entrant and two nonentrants. The score vectors $Q=(1,0,0)$ and $Q=(2,0,0)$ produce the same ranking, but the entrant losses are $\log(1+2e^{-1})$ and $\log(1+2e^{-2})$, respectively. A common shift leaves the softmax unchanged; positive rescaling at a fixed temperature and general monotone transformations do not.

Relative profile thresholds also depend on the objective function. Multiplying the objective by a positive constant does not change relative deterioration. Adding a parameter-independent constant preserves the location of the minimum but changes $|L_{\min}|$ and the corresponding relative ratio. Results for the 603 evaluated points are always defined using the fixed baseline objective; support counts are not compared across different criteria. Table~\ref{tab:app-native-computation} reports the original computational settings.

\subsection{Parameter Bounds and Numerical Checks}

\begin{table}[htbp]
\centering
\caption{Original structural-model computation}
\label{tab:app-native-computation}
\small
\begin{tabular}{ll}
\toprule
Item & Original setting \\
\midrule
Optimizer & Bounded L-BFGS-B \\
Point-estimate starting values & 50 \\
\texttt{maxiter} & 800 \\
\texttt{ftol} & $10^{-12}$ \\
\texttt{gtol} & $10^{-9}$ \\
$\theta$ point-estimate bounds & $[\log(0.05),4]^7$ \\
$\beta$ point-estimate bounds & $[0,10]^2$ \\
$\widehat\beta_C$ & 2.078840 \\
$\widehat\beta_E$ & 0.782621 \\
\bottomrule
\end{tabular}
\end{table}

The structural allocation model is estimated by bounded L-BFGS-B, with multiple starting values and gradient checks used to assess numerical stability. The baseline structural estimation minimizes the same structural objective from 50 starting values, with a maximum of 800 iterations and stopping tolerances of $\texttt{ftol}=10^{-12}$ and $\texttt{gtol}=10^{-9}$. The lowest-loss solution from this search serves as the baseline for the subsequent analysis. All 50 runs return finite results and yield three distinct numerical solutions; 45 objective values lie within the prespecified tolerance of the lowest recorded value. The projected-gradient norm at the baseline solution is $1.575\times10^{-6}$, below the prespecified threshold of $10^{-3}$, and all parameters lie in the interior of the search bounds. Across the stated parameter range and optimization settings, the multiple-start and gradient checks show that the baseline loss and its associated parameters are numerically reproducible, providing a computational reference for the profile and allocation-stability analyses.

The baseline estimation's success, convergence, and projected-gradient checks all pass. The profile-search procedure reoptimizes the remaining parameters at each profile coordinate, and the one-dimensional extended searches expand the relevant bounds to 20. Numerical convergence does not establish global optimality or model identification.

\subsection{Two Fit-Tolerance Scales}

The raw-total-loss deterioration ratio is
\begin{equation}
d(L)=\frac{L-L^*}{|L^*|}.
\tag{B.2}\label{eq:raw-loss-deterioration}
\end{equation}
The share of fit improvement sacrificed is $r(L)$ in Equation~\eqref{eq:fit-sacrifice}. The lowest recorded baseline loss is 763.513744, and the within-set equal-score benchmark loss is 803.251378. Values of $d(L)$ equal to 0.01, 0.02, and 0.05 correspond to $r(L)$ values of 0.192, 0.384, and 0.961, respectively. A 5\% raw-total-loss band therefore permits about 96.1\% of the original fit improvement to be forfeited. Retaining at least 95\% of the fit improvement instead corresponds to $r(L)\leq0.05$.

\setcounter{table}{0}
\renewcommand{\thetable}{C.\arabic{table}}
\setcounter{figure}{0}
\renewcommand{\thefigure}{C.\arabic{figure}}

\section{One-Dimensional Profile Details}
\label{app:one-dimensional-profiles}

To assess how strongly observed Dibao entry constrains the structural parameters, we compare parameter combinations with similar fit rather than rely on a single numerical optimum. For a given pair of food and education gain coefficients, $\beta=(\beta_C,\beta_E)$, the seven household-component parameters $\theta$ are reoptimized to obtain the corresponding profile loss. The resulting profiles show how far the gain coefficients can vary while adjustments in the household component preserve a similar fit, and whether different parameter combinations generate similar household allocations. Following the relative-loss criterion used in the original estimation, define the profile loss and its specified 5\% raw-loss-tolerance region as
\begin{equation}
L_p(\beta)=\min_{\theta}L(\theta,\beta),
\qquad
\widehat\beta\in\arg\min_{\beta}L_p(\beta),
\tag{C.1}
\end{equation}
and
\begin{equation}
B_{\delta}
=\left\{\beta:
\frac{L_p(\beta)-L_p(\widehat\beta)}{|L_p(\widehat\beta)|}
\leq\delta\right\},
\qquad \delta=0.05.
\tag{C.2}
\end{equation}

For a fixed objective function, this criterion compares the profile loss after reoptimizing the household component with the baseline minimum loss, providing a common standard for evaluating the range of gain coefficients. The specified loss-tolerance region is characterized using the parameter coordinates already evaluated and stored together with their numerical optimization results; the finite grid is not treated as the entire feasible parameter space. A proportional increase in raw total loss and the share of fit improvement forfeited relative to the within-set equal-score benchmark are distinct scales. The former measures deterioration from the minimum objective value, whereas the latter measures how much of the original fit improvement is relinquished. Because parameter-independent constants in the raw loss affect the first ratio, 5\% raw-loss-tolerance regions cannot be compared directly across different loss functions, and the resulting parameter sets are not statistical confidence regions (Appendix B). We retain the original total-loss criterion to describe the geometry of the stored parameter profiles. Section 6.4 and Appendix L instead use loss normalized relative to the equal-score benchmark to examine the relationship between fit tolerance and simulated-list stability. This additional comparison links parameter variation to allocation outcomes and distinguishes substitution under a looser fit requirement from allocation stability under a stricter one.

The profile-search analysis examines how much reoptimization of the remaining parameters can preserve allocation fit as the food and education gain coefficients change. The one-dimensional food profile fixes $\beta_C$ and reoptimizes $\beta_E$ together with the seven household-component parameters. The education profile fixes $\beta_E$ and reoptimizes $\beta_C$ and the seven household-component parameters. The joint profile fixes both gain coefficients and reoptimizes only the household component. The one-dimensional profiles and the additional two-dimensional coordinates pass the projected-gradient checks for the free parameters, while the base two-dimensional grid retains the archived baseline records. Allowing the household component to adjust with the gain coefficients makes profile loss informative about parameter compensation across allocation components and about the fit cost of changing a gain coefficient. The subsequent analysis compares household-characteristic relationships, gain parameters, and the corresponding complete household rankings to evaluate how strongly observed entry constrains each level of the allocation results.

Tables~\ref{tab:app-food-profile} and \ref{tab:app-education-profile} retain the archived profile coordinates and threshold classifications. Percentage deterioration is $[L_p-L_{\min}]/|L_{\min}|\times100\%$. The food-consumption profile deteriorates by about 3.873\% at 20 and remains open at that boundary. The final supported education grid point is 5.25; deterioration at the next point, 5.40, is approximately 5.001\%. This endpoint is conditional on the other search bounds. In the upper range of the education coefficient, the reoptimized food coefficient reaches 20, so the table does not establish closure over an unbounded parameter space.

\begin{longtable}{rrrrrr}
\caption{Food-consumption relative-value profile}
\label{tab:app-food-profile}\\
\toprule
$\beta_C$ & Profiled $\beta_E$ & Deterioration (\%) & $\leq1\%$ & $\leq2\%$ & $\leq5\%$ \\
\midrule
\endfirsthead
\multicolumn{6}{l}{\textit{Table C.1 continued}}\\
\toprule
$\beta_C$ & Profiled $\beta_E$ & Deterioration (\%) & $\leq1\%$ & $\leq2\%$ & $\leq5\%$ \\
\midrule
\endhead
\midrule
\multicolumn{6}{r}{\textit{Continued on next page}}\\
\endfoot
\bottomrule
\endlastfoot
0.00 & 0.000 & 0.227 & Yes & Yes & Yes \\
0.25 & 0.000 & 0.160 & Yes & Yes & Yes \\
0.50 & 0.034 & 0.133 & Yes & Yes & Yes \\
0.75 & 0.149 & 0.124 & Yes & Yes & Yes \\
1.00 & 0.279 & 0.110 & Yes & Yes & Yes \\
1.25 & 0.413 & 0.087 & Yes & Yes & Yes \\
1.50 & 0.533 & 0.056 & Yes & Yes & Yes \\
1.75 & 0.643 & 0.024 & Yes & Yes & Yes \\
2.00 & 0.749 & 0.002 & Yes & Yes & Yes \\
2.25 & 0.854 & 0.011 & Yes & Yes & Yes \\
2.50 & 0.953 & 0.079 & Yes & Yes & Yes \\
2.75 & 1.040 & 0.244 & Yes & Yes & Yes \\
3.00 & 1.106 & 0.551 & Yes & Yes & Yes \\
3.25 & 1.152 & 0.986 & Yes & Yes & Yes \\
3.50 & 1.181 & 1.442 & No & Yes & Yes \\
3.75 & 1.188 & 1.864 & No & Yes & Yes \\
4.00 & 1.179 & 2.234 & No & No & Yes \\
4.25 & 1.153 & 2.535 & No & No & Yes \\
4.50 & 1.105 & 2.760 & No & No & Yes \\
4.75 & 1.031 & 2.917 & No & No & Yes \\
5.00 & 0.929 & 3.021 & No & No & Yes \\
5.25 & 0.807 & 3.092 & No & No & Yes \\
5.50 & 0.660 & 3.142 & No & No & Yes \\
5.75 & 0.496 & 3.183 & No & No & Yes \\
6.00 & 0.327 & 3.217 & No & No & Yes \\
6.50 & 0.000 & 3.272 & No & No & Yes \\
7.00 & 0.000 & 3.318 & No & No & Yes \\
7.50 & 0.000 & 3.360 & No & No & Yes \\
8.00 & 0.000 & 3.398 & No & No & Yes \\
8.50 & 0.000 & 3.433 & No & No & Yes \\
9.00 & 0.000 & 3.466 & No & No & Yes \\
9.50 & 0.000 & 3.497 & No & No & Yes \\
10.00 & 0.000 & 3.525 & No & No & Yes \\
11.00 & 0.000 & 3.577 & No & No & Yes \\
12.00 & 0.000 & 3.623 & No & No & Yes \\
13.00 & 0.000 & 3.664 & No & No & Yes \\
14.00 & 0.000 & 3.702 & No & No & Yes \\
15.00 & 0.000 & 3.736 & No & No & Yes \\
16.00 & 0.000 & 3.767 & No & No & Yes \\
17.00 & 0.000 & 3.796 & No & No & Yes \\
18.00 & 0.000 & 3.823 & No & No & Yes \\
19.00 & 0.000 & 3.849 & No & No & Yes \\
20.00 & 0.000 & 3.873 & No & No & Yes \\
\end{longtable}

\begin{longtable}{rrrrrr}
\caption{Education relative-value profile}
\label{tab:app-education-profile}\\
\toprule
$\beta_E$ & Profiled $\beta_C$ & Deterioration (\%) & $\leq1\%$ & $\leq2\%$ & $\leq5\%$ \\
\midrule
\endfirsthead
\multicolumn{6}{l}{\textit{Table C.2 continued}}\\
\toprule
$\beta_E$ & Profiled $\beta_C$ & Deterioration (\%) & $\leq1\%$ & $\leq2\%$ & $\leq5\%$ \\
\midrule
\endhead
\midrule
\multicolumn{6}{r}{\textit{Continued on next page}}\\
\endfoot
\bottomrule
\endlastfoot
0.00 & 0.441 & 0.136 & Yes & Yes & Yes \\
0.15 & 0.780 & 0.123 & Yes & Yes & Yes \\
0.30 & 1.095 & 0.105 & Yes & Yes & Yes \\
0.45 & 1.400 & 0.074 & Yes & Yes & Yes \\
0.60 & 1.708 & 0.033 & Yes & Yes & Yes \\
0.75 & 2.014 & 0.001 & Yes & Yes & Yes \\
0.90 & 2.315 & 0.026 & Yes & Yes & Yes \\
1.05 & 2.646 & 0.206 & Yes & Yes & Yes \\
1.20 & 3.062 & 0.913 & Yes & Yes & Yes \\
1.35 & 3.676 & 2.346 & No & No & Yes \\
1.50 & 4.833 & 3.509 & No & No & Yes \\
1.65 & 7.129 & 3.920 & No & No & Yes \\
1.80 & 10.867 & 4.077 & No & No & Yes \\
1.95 & 14.802 & 4.157 & No & No & Yes \\
2.10 & 18.152 & 4.213 & No & No & Yes \\
2.25 & 20.000 & 4.258 & No & No & Yes \\
2.40 & 20.000 & 4.304 & No & No & Yes \\
2.55 & 20.000 & 4.354 & No & No & Yes \\
2.70 & 20.000 & 4.408 & No & No & Yes \\
2.85 & 20.000 & 4.466 & No & No & Yes \\
3.00 & 20.000 & 4.528 & No & No & Yes \\
3.15 & 20.000 & 4.594 & No & No & Yes \\
3.30 & 20.000 & 4.662 & No & No & Yes \\
3.45 & 20.000 & 4.731 & No & No & Yes \\
3.60 & 20.000 & 4.800 & No & No & Yes \\
3.75 & 20.000 & 4.865 & No & No & Yes \\
3.90 & 20.000 & 4.922 & No & No & Yes \\
4.05 & 20.000 & 4.940 & No & No & Yes \\
4.20 & 20.000 & 4.945 & No & No & Yes \\
4.35 & 20.000 & 4.951 & No & No & Yes \\
4.50 & 20.000 & 4.957 & No & No & Yes \\
4.65 & 20.000 & 4.963 & No & No & Yes \\
4.80 & 20.000 & 4.970 & No & No & Yes \\
4.95 & 20.000 & 4.977 & No & No & Yes \\
5.10 & 20.000 & 4.985 & No & No & Yes \\
5.25 & 20.000 & 4.992 & No & No & Yes \\
5.40 & 20.000 & 5.001 & No & No & No \\
5.55 & 20.000 & 5.009 & No & No & No \\
5.70 & 20.000 & 5.018 & No & No & No \\
5.85 & 20.000 & 5.027 & No & No & No \\
6.00 & 20.000 & 5.036 & No & No & No \\
\end{longtable}

\setcounter{table}{0}
\renewcommand{\thetable}{D.\arabic{table}}
\setcounter{figure}{0}
\renewcommand{\thefigure}{D.\arabic{figure}}

\section{Joint Geometry and Household-Component Directions}
\label{app:joint-geometry}

The joint profile consists of 1,681 base coordinates on $[0,6]\times[0,6]$ with a step size of 0.15 and 87 deduplicated extended coordinates, for a total of 1,768 evaluated parameter points. Of these, 603 satisfy the 5\% relative-loss criterion under the baseline objective. The seven household-component coefficients are reoptimized at every evaluated coordinate. Across these 603 evaluated coordinates, the income-rank coefficient is negative at every point, the household-head education coefficient is negative at 599 points, and the elderly-member coefficient is positive at 591 points. Table~\ref{tab:app-sign-reversals} reports all four records in which the education coefficient becomes positive and all 12 records in which the elderly-member coefficient becomes negative. Because $(\beta_C,\beta_E)=(15,4)$ and $(20,4)$ each generate reversals in both coefficients, the 16 records correspond to 14 distinct coordinates. The point $(6,1.95)$ belongs to the base grid, whereas $(6,2.00)$ is an off-grid extended coordinate. The other 12 distinct reversal coordinates lie in the extended region where the food-gain coefficient equals 15 or 20. The three household-component directions therefore recur at high rates along the evaluated conditional-optimum paths; these frequencies do not identify their signs throughout the continuous 5\% raw-loss-tolerance region.

\begin{figure}[htbp]
\centering
\includegraphics[width=\textwidth]{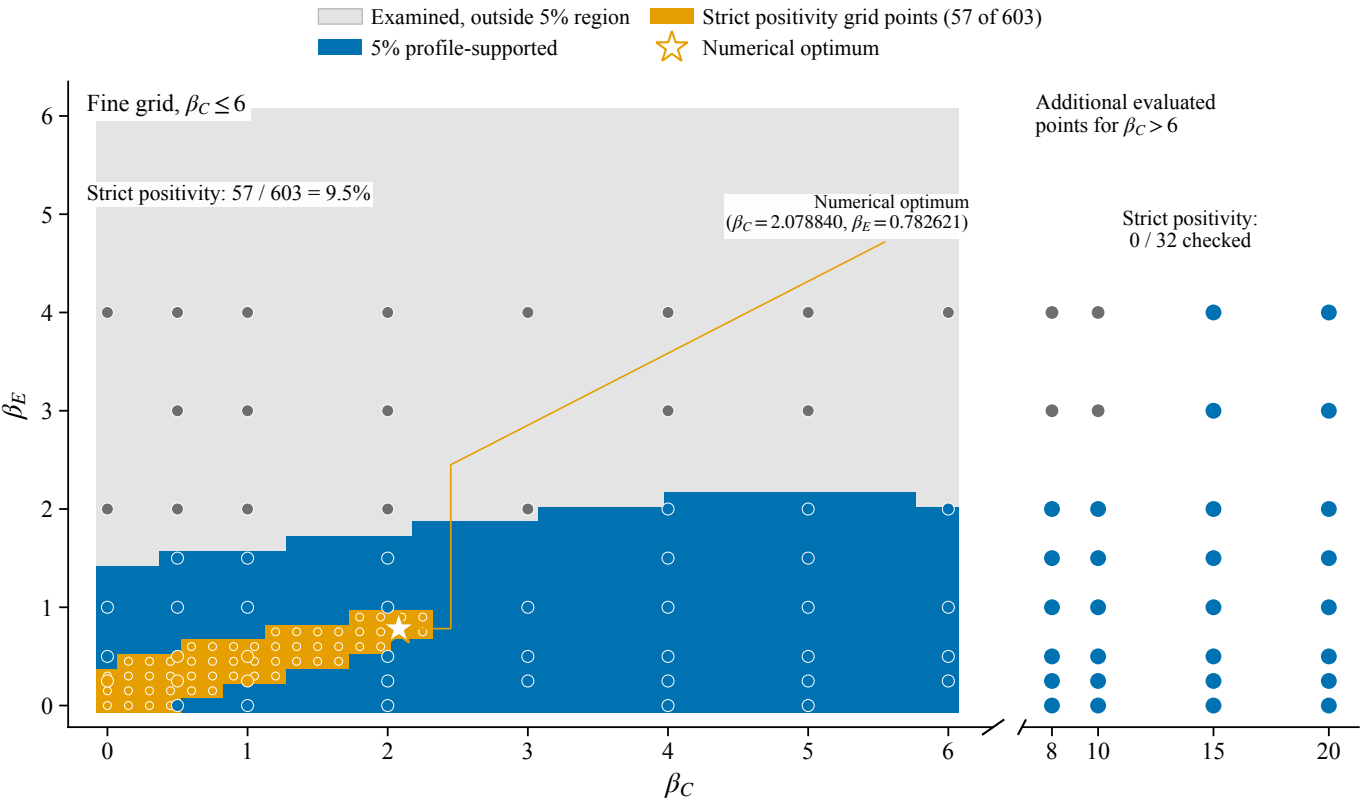}
\caption{Geometry of outcome-value parameters supported by the joint profile}
\label{fig:app-joint-geometry}
\parbox{0.96\textwidth}{\footnotesize\textit{Note:} Colored cells are coordinates actually evaluated on the grid with step size 0.15; no interpolation is used. Circles denote additional off-grid coordinates and are colored by category. The right panel contains the additional evaluated points with $\beta_C>6$ and uses a different horizontal scale. Orange denotes coordinates satisfying strict positivity for all 3,102 households, and the star marks the numerical optimum.}
\end{figure}

\begin{table}[htbp]
\centering
\caption{Stability of conditional household-component relationships}
\label{tab:app-direction-stability}
\small
\begin{tabular}{lccc}
\toprule
Household characteristic & 5\% raw-loss-tolerance joint grid & Direction & Positive-multiplier subset \\
\midrule
Income rank & 603/603 (100\%) & Negative & 57/57 \\
Education of household head & 599/603 (99.34\%) & Negative & 57/57 \\
Elderly member & 591/603 (98.01\%) & Positive & 57/57 \\
\bottomrule
\end{tabular}
\parbox{0.96\textwidth}{\footnotesize\textit{Note:} The full grid contains the 603 evaluated parameter coordinates that satisfy the specified 5\% raw-total-loss criterion under the baseline objective. The strict-positivity subset contains 57 coordinates for which $G_i(\beta)>0$ for every household in the structural sample. Both sets use the same baseline objective and fixed original gain predictions. Directional consistency rates describe the distribution of coefficients over evaluated coordinates; they are neither statistical significance levels nor population probabilities.}
\end{table}

\begin{table}[htbp]
\centering
\caption{Sign reversals within the baseline-support grid}
\label{tab:app-sign-reversals}
\small
\begin{tabular}{lrrrr}
\toprule
Relationship & $\beta_C$ & $\beta_E$ & Coefficient & Deterioration (\%) \\
\midrule
Education of household head & 6.00 & 1.95 & 0.082668 & 4.871 \\
Education of household head & 6.00 & 2.00 & 0.163050 & 4.935 \\
Education of household head & 15.00 & 4.00 & 1.353258 & 4.984 \\
Education of household head & 20.00 & 4.00 & 1.105050 & 4.939 \\
Elderly member & 15.00 & 0.00 & -0.196243 & 3.736 \\
Elderly member & 15.00 & 0.25 & -0.154916 & 3.762 \\
Elderly member & 15.00 & 0.50 & -0.111457 & 3.794 \\
Elderly member & 15.00 & 1.00 & -0.017529 & 3.877 \\
Elderly member & 15.00 & 4.00 & -2.995732 & 4.984 \\
Elderly member & 20.00 & 0.00 & -0.329507 & 3.873 \\
Elderly member & 20.00 & 0.25 & -0.301099 & 3.897 \\
Elderly member & 20.00 & 0.50 & -0.271678 & 3.924 \\
Elderly member & 20.00 & 1.00 & -0.209627 & 3.990 \\
Elderly member & 20.00 & 1.50 & -0.140962 & 4.076 \\
Elderly member & 20.00 & 2.00 & -0.068169 & 4.189 \\
Elderly member & 20.00 & 4.00 & -2.995732 & 4.939 \\
\bottomrule
\end{tabular}
\end{table}

\setcounter{table}{0}
\renewcommand{\thetable}{E.\arabic{table}}
\setcounter{figure}{0}
\renewcommand{\thefigure}{E.\arabic{figure}}

\section{Strict Positivity}
\label{app:strict-positivity}

\subsection{The Positivity Condition and the Two Channels of Household Characteristics}

The relationship between the household component and the complete allocation score depends on the sign of the gain multiplier. Under the score representation in Section 3.1, the household component is always positive, while $G_i(\beta)$ varies with the household's predicted policy gains and the corresponding coefficients. Holding the predicted gains and their coefficients fixed,
\begin{equation}
Q_i=\omega_iG_i,
\qquad
\frac{\partial Q_i}{\partial\omega_i}=G_i(\beta).
\tag{E.1}
\end{equation}
When $G_i(\beta)>0$, a higher household component implies a higher complete allocation score, giving the directions of the household-component coefficients a clear allocation interpretation conditional on the gain multiplier. We therefore distinguish the previously evaluated solutions in the specified raw-loss-tolerance region that satisfy strict positivity and examine the directions of household-characteristic coefficients and the stability of complete allocations within that subset. Strict positivity requires the gain multiplier to be positive for every household in the structural sample. It is applied after estimation as a filter and is not imposed as an additional restriction in the baseline or sensitivity optimizations. The specific opposite-direction constraint searches in Sections 5.3 and 6.1 instead include positivity as a feasibility requirement. The positive-multiplier subset links the directional stability of household-component relationships to the corresponding complete allocation rankings, providing a common basis for comparing the economic interpretation of different parameter combinations.

Household characteristics affect the complete allocation score through both the household component and predicted policy gains. Income rank, household-head education, and household demographic composition enter the exponential household component $\omega(X_i;\theta)$. The same characteristics also enter the predictions of food-consumption and education-spending gains and therefore affect $G_i(\beta)$. A household-component coefficient thus describes the conditional relationship between a characteristic and the baseline allocation component when predicted gains are held fixed. Evaluating the characteristic's overall relationship with the complete score also requires accounting for the corresponding change in predicted gains. For a continuous characteristic, the two channels can be assessed from the rates at which the household component and gain multiplier change with that characteristic. For binary characteristics, such as the presence of children or elderly members, the complete score must be recomputed at each characteristic value. Together, the two channels determine households' relative allocation positions and allow changes in household-component and gain coefficients to offset one another in the complete score. This distinction supports separate analysis of the stability of conditional household relationships and of how those relationships combine with predicted gains to form household rankings and fixed-quota simulated lists.

\subsection{Evaluated Strict-Positivity Coordinates}

Of the 603 evaluated coordinates satisfying the specified 5\% raw-loss-tolerance criterion, 57 satisfy $G_i=1+\beta_C\widehat\tau_{iC}+\beta_E\widehat\tau_{iE}>0$ for all 3,102 households. The coordinates span $\beta_C\in[0,2.25]$ and $\beta_E\in[0,0.90]$, and all three reported coefficient directions hold at all 57 points. Table~\ref{tab:app-positive-coordinates} records every coordinate and the minimum $G_i$ across households. This is an interpretive subset of the evaluated grid, not a formal identified set or a new estimation that imposes positivity.

\begin{table}[htbp]
\centering
\caption{Coordinates satisfying strict positivity}
\label{tab:app-positive-coordinates}
\footnotesize
\setlength{\tabcolsep}{4.5pt}
\begin{tabular}{rrrrrrrrr}
\toprule
$\beta_C$ & $\beta_E$ & Minimum $G_i$ & $\beta_C$ & $\beta_E$ & Minimum $G_i$ & $\beta_C$ & $\beta_E$ & Minimum $G_i$ \\
\midrule
0.00 & 0.00 & 1.000000 & 0.60 & 0.30 & 0.680626 & 1.35 & 0.45 & 0.403164 \\
0.00 & 0.15 & 0.645375 & 0.60 & 0.45 & 0.359983 & 1.35 & 0.60 & 0.374605 \\
0.00 & 0.25 & 0.408959 & 0.60 & 0.60 & 0.038632 & 1.35 & 0.75 & 0.122327 \\
0.00 & 0.30 & 0.290750 & 0.75 & 0.15 & 0.303085 & 1.50 & 0.45 & 0.265723 \\
0.15 & 0.00 & 0.678795 & 0.75 & 0.30 & 0.677795 & 1.50 & 0.60 & 0.355591 \\
0.15 & 0.15 & 0.759658 & 0.75 & 0.45 & 0.440992 & 1.50 & 0.75 & 0.201565 \\
0.15 & 0.30 & 0.415261 & 0.75 & 0.60 & 0.119641 & 1.65 & 0.45 & 0.115762 \\
0.15 & 0.45 & 0.060636 & 0.90 & 0.30 & 0.602109 & 1.65 & 0.60 & 0.279905 \\
0.30 & 0.00 & 0.357589 & 0.90 & 0.45 & 0.520939 & 1.65 & 0.75 & 0.220577 \\
0.30 & 0.15 & 0.840313 & 0.90 & 0.60 & 0.200650 & 1.80 & 0.60 & 0.204219 \\
0.30 & 0.30 & 0.519316 & 1.00 & 0.25 & 0.373817 & 1.80 & 0.75 & 0.211294 \\
0.30 & 0.45 & 0.182843 & 1.00 & 0.50 & 0.467710 & 1.80 & 0.90 & 0.041878 \\
0.45 & 0.00 & 0.036384 & 1.05 & 0.30 & 0.460495 & 1.95 & 0.60 & 0.070951 \\
0.45 & 0.15 & 0.801055 & 1.05 & 0.45 & 0.528633 & 1.95 & 0.75 & 0.156646 \\
0.45 & 0.30 & 0.600325 & 1.05 & 0.60 & 0.281659 & 1.95 & 0.90 & 0.066549 \\
0.45 & 0.45 & 0.278974 & 1.20 & 0.30 & 0.248580 & 2.10 & 0.75 & 0.080960 \\
0.50 & 0.25 & 0.733855 & 1.20 & 0.45 & 0.478850 & 2.10 & 0.90 & 0.057266 \\
0.50 & 0.50 & 0.198860 & 1.20 & 0.60 & 0.361252 & 2.25 & 0.75 & 0.005274 \\
0.60 & 0.15 & 0.624290 & 1.20 & 0.75 & 0.041318 & 2.25 & 0.90 & 0.033386 \\
\bottomrule
\end{tabular}
\end{table}

\setcounter{table}{0}
\renewcommand{\thetable}{F.\arabic{table}}
\setcounter{figure}{0}
\renewcommand{\thefigure}{F.\arabic{figure}}

\section{Fixed-Quota Allocation Sensitivity}
\label{app:fixed-quota-sensitivity}

For each parameterization, households are ranked by $Q_i$ within each comparison set, and the top $K_m$ households are selected so that the observed number of entrants is preserved. Overlap with the observed entrant list is the size of the intersection divided by 265. Allocation status that remains unchanged across parameterizations includes households that are always selected and those that are never selected, with all 3,102 households as the denominator. Tables~\ref{tab:app-observed-overlap} and \ref{tab:app-allocation-status} report these two distinct measures.

The list comparisons in this appendix use the 603 evaluated parameter combinations that satisfy the 5\% raw-total-loss criterion. Quotas remain fixed in every comparison set.

\begin{figure}[htbp]
\centering
\includegraphics[width=\textwidth]{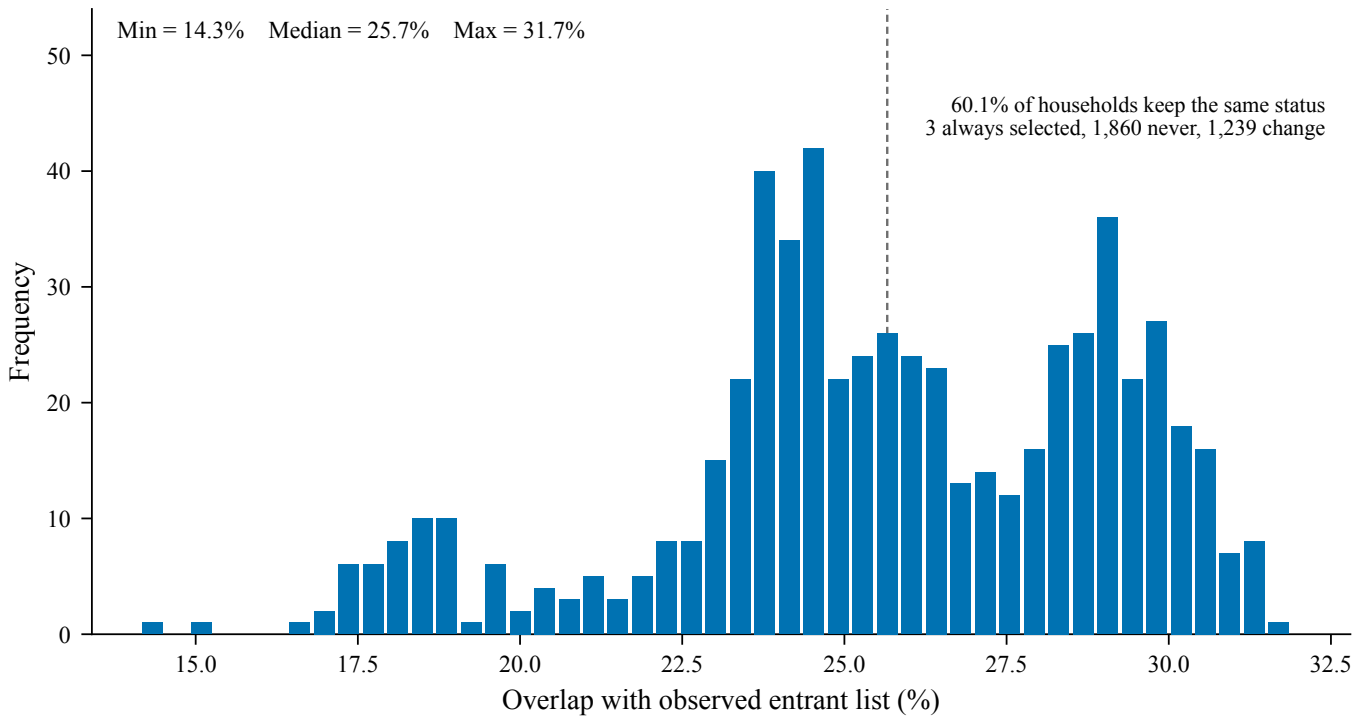}
\caption{Allocation sensitivity across profile-supported parameterizations}
\label{fig:app-allocation-sensitivity}
\parbox{0.96\textwidth}{\footnotesize\textit{Note:} Each bar gives the number of parameterizations, among the 603 satisfying the baseline profile's 5\% criterion, with the indicated overlap with the 265 observed entrants. The dashed line marks the median overlap of 25.7\%. The value 60.1\% is not the height of a bar; it is the share of all 3,102 households whose selection status remains unchanged across the 603 parameterizations.}
\end{figure}

\begin{table}[htbp]
\centering
\caption{Overlap with observed entrants}
\label{tab:app-observed-overlap}
\small
\begin{tabular}{lr}
\toprule
Statistic & Overlap with observed entrants \\
\midrule
Minimum & 14.3\% \\
10th percentile & 20.8\% \\
25th percentile & 23.8\% \\
Median & 25.7\% \\
75th percentile & 28.7\% \\
90th percentile & 29.8\% \\
Maximum & 31.7\% \\
\bottomrule
\end{tabular}
\end{table}

\begin{table}[htbp]
\centering
\caption{Allocation status across 603 parameterizations}
\label{tab:app-allocation-status}
\small
\begin{tabular}{lr}
\toprule
Status & Households \\
\midrule
Always selected & 3 \\
Never selected & 1,860 \\
Status changes & 1,239 \\
Total & 3,102 \\
\bottomrule
\end{tabular}
\parbox{0.96\textwidth}{\footnotesize\textit{Note:} Households with unchanged status account for 60.1\% of the full sample. The model-generated lists do not have an independently established welfare ranking, so differences in overlap cannot be interpreted directly as targeting error or welfare loss.}
\end{table}

\setcounter{table}{0}
\renewcommand{\thetable}{G.\arabic{table}}
\setcounter{figure}{0}
\renewcommand{\thefigure}{G.\arabic{figure}}

\section{Prespecified Loss Sensitivity and County Reweighting}
\label{app:loss-sensitivity}

\subsection{Fixed Protocol}

The protocol fixed five specifications and their comparison metrics before the sensitivity models were fitted. Every specification retains the 3,102 households, 265 entrants, 125 comparison sets, 107 counties, seven covariates, the original fixed gain predictions, and the original point-estimate bounds. No intercept or positivity constraint is added. Survey weights are normalized once to have mean one over the full structural sample. Define
\begin{equation}
W_m^+=\sum_{D_i=1}s_i,
\qquad
W_m^-=\sum_{D_i=0}s_i.
\tag{G.1}\label{eq:G-1}
\end{equation}
The equal-weight-across-sets softmax specification uses $\sum_m L_m/W_m^+$. The equal-weight-across-sets pairwise criterion compares entrant $i$ with nonentrant $j$:
\begin{equation}
L^{\mathrm{pair}}
=\sum_m\sum_{i:D_i=1}\sum_{j:D_j=0}
\frac{s_i}{W_m^+}\frac{s_j}{W_m^-}
\log\!\left(1+e^{-(Q_i-Q_j)}\right).
\tag{G.2}\label{eq:G-2}
\end{equation}
Pair weights sum to one within each comparison set, so larger sets do not receive greater total weight solely because they contain more pairs. The temperature specifications use
\begin{equation}
p_i(T)=\frac{e^{Q_i/T}}{\sum_{j\in m}e^{Q_j/T}},
\qquad T\in\{0.5,1,2\}.
\tag{G.3}\label{eq:G-3}
\end{equation}
These variations assess sensitivity to objective weights, ranking loss, and score temperature; they are not mathematically equivalent estimators.

Numerical checks divide the full objective by the fixed number of 125 comparison sets, whereas the reported total objective values are not divided by this number. Full-sample estimation uses 50 common starting values and random seed 20260928. County reweighting uses 30 common vectors of $\operatorname{Exp}(1)$ multipliers, normalized to have county-level mean one, with random seed 20260929. The same perturbation is applied to all five specifications and multiplies the entire comparison-set loss. Each specification and perturbation uses 20 predetermined starting values, including the specification's full-sample solution, the original baseline solution, and fixed starts; some starting values are repeated.

Optimization uses bounded L-BFGS-B with \texttt{maxiter}$=800$, \texttt{maxfun}$=16000$, \texttt{ftol}$=10^{-12}$, and \texttt{gtol}$=10^{-9}$. Numerical validity requires the $L_2$ norm of the projected gradient on the average-loss scale to be no greater than $10^{-5}$. All starting-value records are retained. When needed, the protocol permits one numerical refinement from a candidate solution under the same bounds, without changing the seed, bounds, or criterion in response to the results. All five full-sample solutions and all 150 reweighted solutions pass these checks. Numerical validity does not establish global optimality.

\subsection{Full-Sample Coefficients and Common Covariate Changes}

Table~\ref{tab:app-full-sample-coefficients} reports raw coefficients in the specified parameterization. Coefficient ratios across criteria are not interpreted as effect ratios or welfare-value ratios. Table~\ref{tab:app-common-covariate-changes} instead reports a common set of covariate changes within each specification. Table~\ref{tab:app-positivity-concordance} reports strict positivity, in-sample pairwise concordance, and the number of parameters at a bound.

\begin{table}[htbp]
\centering
\caption{Full-sample coefficients}
\label{tab:app-full-sample-coefficients}
\footnotesize
\resizebox{\textwidth}{!}{%
\begin{tabular}{lrrrrr}
\toprule
Parameter & Baseline & Set-equal softmax & Set-equal pairwise & $T=0.5$ & $T=2$ \\
\midrule
Income rank & -1.376683 & -1.268744 & -1.108942 & -0.725208 & -1.898893 \\
Household size/10 & 0.045977 & -0.477754 & -0.284897 & 0.203097 & 0.036626 \\
Child present & -0.085099 & -0.036025 & -0.022789 & -0.068480 & -0.138790 \\
Elderly member & 0.870719 & 0.901409 & 0.804409 & 0.507184 & 1.105549 \\
Female household head & 0.097326 & 0.017183 & -0.040245 & 0.041275 & 0.130591 \\
Age of household head/100 & 1.404111 & 2.003165 & 1.997888 & 0.785194 & 2.316029 \\
Education code/9 & -1.511796 & -1.730468 & -1.513011 & -0.850328 & -1.916358 \\
$\beta_C$ & 2.078840 & 2.170833 & 2.044261 & 1.391693 & 2.497428 \\
$\beta_E$ & 0.782621 & 0.897657 & 0.831166 & 0.508433 & 0.939742 \\
\bottomrule
\end{tabular}%
}
\end{table}

\begin{table}[htbp]
\centering
\caption{Household-component changes under common covariate changes}
\label{tab:app-common-covariate-changes}
\small
\begin{tabular}{lrrr}
\toprule
Specification & Income rank $+0.1$ & Education code $+1$ & Elderly indicator $0\to1$ \\
\midrule
Baseline $T=1$ & -12.86\% & -15.46\% & +138.86\% \\
Set-equal softmax & -11.92\% & -17.49\% & +146.31\% \\
Set-equal pairwise & -10.50\% & -15.47\% & +123.54\% \\
$T=0.5$ & -7.00\% & -9.02\% & +66.06\% \\
$T=2$ & -17.29\% & -19.18\% & +202.09\% \\
\bottomrule
\end{tabular}
\parbox{0.96\textwidth}{\footnotesize\textit{Note:} Reported values equal $100[\exp(\theta_k\Delta X_k)-1]$. All other inputs, including generated gains, are held fixed. These are conditional changes in $\omega$.}
\end{table}

\begin{table}[htbp]
\centering
\caption{Full-sample positivity and pairwise concordance}
\label{tab:app-positivity-concordance}
\small
\begin{tabular}{lrrrr}
\toprule
Specification & Share with $G_i>0$ & Minimum $G_i$ & Pairwise concordance & Parameters at bounds \\
\midrule
Baseline $T=1$ & 100.0000\% & 0.097750 & 0.672318 & 0 \\
Set-equal softmax & 100.0000\% & 0.054998 & 0.682405 & 0 \\
Set-equal pairwise & 100.0000\% & 0.122877 & 0.681857 & 0 \\
$T=0.5$ & 100.0000\% & 0.393078 & 0.679812 & 0 \\
$T=2$ & 99.9033\% & -0.084011 & 0.670909 & 0 \\
\bottomrule
\end{tabular}
\parbox{0.96\textwidth}{\footnotesize\textit{Note:} Pairwise concordance uses normalized survey weights for entrants and nonentrants, equal weighting across comparison sets, and half weight for ties. These measures describe in-sample performance. Under $T=2$, three households fail strict positivity.}
\end{table}

\begin{table}[htbp]
\centering
\caption{Full-sample directions, rankings, and lists}
\label{tab:app-full-sample-lists}
\footnotesize
\resizebox{\textwidth}{!}{%
\begin{tabular}{lccccc}
\toprule
Specification & Income/education/elderly & $Q$ correlation & $\omega$ correlation & Baseline-list overlap & Observed-entrant overlap \\
\midrule
Baseline $T=1$ & $-/-/+$ & 1.000 & 1.000 & 100.0\% & 29.1\% \\
Set-equal softmax & $-/-/+$ & 0.963 & 0.986 & 75.8\% & 30.9\% \\
Set-equal pairwise & $-/-/+$ & 0.953 & 0.983 & 74.0\% & 32.1\% \\
$T=0.5$ & $-/-/+$ & 0.984 & 0.990 & 90.6\% & 27.9\% \\
$T=2$ & $-/-/+$ & 0.988 & 0.995 & 89.1\% & 30.6\% \\
\bottomrule
\end{tabular}%
}
\parbox{0.96\textwidth}{\footnotesize\textit{Note:} The three signs refer, in order, to the household-component coefficients on income rank, household-head education, and elderly-member status. Correlations are equal-weighted means of within-set Spearman correlations relative to the fixed baseline model. Each model list retains the observed entrant count $K_m$ in every comparison set, and each overlap rate is the relevant list intersection divided by 265. Baseline-list overlap and observed-entrant overlap use different comparison lists. All five full-sample estimates pass the prespecified numerical checks.}
\end{table}

\subsection{Reweighting Distributions}

Tables~\ref{tab:app-reweight-score-correlation} and \ref{tab:app-reweight-list-overlap} compare every reweighted solution with the fixed baseline structural model. Spearman correlations are calculated within comparison sets and then averaged with equal weight across sets. None of the full-sample specifications excludes a set because its scores are constant. Each list retains the original $K_m$, with ties broken lexicographically by household key. The quantiles and extrema summarize 30 perturbations with fixed inputs; they are not confidence intervals that incorporate first-stage reestimation.

\begin{table}[htbp]
\centering
\caption{Within-set score correlations under county reweighting}
\label{tab:app-reweight-score-correlation}
\small
\begin{tabular}{lrrrrr}
\toprule
Specification & Minimum & 5th percentile & Median & 95th percentile & Maximum \\
\midrule
Baseline $T=1$ & 0.813 & 0.838 & 0.914 & 0.949 & 0.958 \\
Set-equal softmax & -0.320 & 0.750 & 0.902 & 0.954 & 0.961 \\
Set-equal pairwise & 0.725 & 0.746 & 0.874 & 0.944 & 0.952 \\
$T=0.5$ & 0.825 & 0.834 & 0.909 & 0.938 & 0.962 \\
$T=2$ & 0.812 & 0.835 & 0.913 & 0.949 & 0.958 \\
\bottomrule
\end{tabular}
\end{table}

\begin{table}[htbp]
\centering
\caption{Model-list overlap under county reweighting}
\label{tab:app-reweight-list-overlap}
\small
\begin{tabular}{lrrrrr}
\toprule
Specification & Minimum & 5th percentile & Median & 95th percentile & Maximum \\
\midrule
Baseline $T=1$ & 58.5\% & 63.5\% & 72.3\% & 78.5\% & 84.5\% \\
Set-equal softmax & 4.2\% & 51.3\% & 67.7\% & 77.6\% & 81.9\% \\
Set-equal pairwise & 43.8\% & 50.2\% & 62.6\% & 76.1\% & 76.6\% \\
$T=0.5$ & 57.7\% & 62.6\% & 70.6\% & 78.9\% & 83.0\% \\
$T=2$ & 60.0\% & 63.9\% & 72.8\% & 78.3\% & 83.0\% \\
\bottomrule
\end{tabular}
\end{table}

\begin{table}[htbp]
\centering
\caption{Directions under common county reweighting}
\label{tab:app-reweight-directions}
\small
\begin{tabular}{lrrrr}
\toprule
Specification & Income negative & Education negative & Elderly positive & Numerically valid \\
\midrule
Baseline $T=1$ & 30/30 & 29/30 & 30/30 & 30/30 \\
Set-equal softmax & 29/30 & 30/30 & 28/30 & 30/30 \\
Set-equal pairwise & 30/30 & 30/30 & 26/30 & 30/30 \\
$T=0.5$ & 30/30 & 30/30 & 30/30 & 30/30 \\
$T=2$ & 30/30 & 28/30 & 30/30 & 30/30 \\
\bottomrule
\end{tabular}
\parbox{0.96\textwidth}{\footnotesize\textit{Note:} All five specifications use the same 30 vectors of county multipliers while holding household records, covariates, comparison sets, and predicted gains fixed. Direction counts describe numerical estimates under the prespecified weight perturbations; they are neither population sign probabilities nor statistical confidence levels.}
\end{table}

\subsection{The Fourteenth Reweighting}

Under the fourteenth perturbation, indexed 13 from zero, the equal-weight-across-sets softmax solution differs sharply from the fixed baseline. Tables~\ref{tab:app-fourteenth-parameters} and \ref{tab:app-fourteenth-diagnostics} report all parameters and diagnostics. Under the prespecified fit without an added positivity constraint, the protocol retains the solution with the lower objective value.

\begin{table}[htbp]
\centering
\caption{Parameters for the fourteenth equal-weight-across-sets softmax reweighting}
\label{tab:app-fourteenth-parameters}
\small
\begin{tabular}{lrl}
\toprule
Parameter & Estimate & Bound status \\
\midrule
Income rank & 2.737793 & Interior \\
Household size/10 & 4.000000 & Upper bound \\
Child present & -2.995732 & Lower bound \\
Elderly member & -1.386371 & Interior \\
Female household head & -2.995732 & Lower bound \\
Age of household head/100 & -2.995732 & Lower bound \\
Education code/9 & -2.995732 & Lower bound \\
$\beta_C$ & 1.755870 & Interior \\
$\beta_E$ & 0.000000 & Lower bound \\
\bottomrule
\end{tabular}
\end{table}

\begin{table}[htbp]
\centering
\caption{Diagnostics for the sharply different solution}
\label{tab:app-fourteenth-diagnostics}
\small
\begin{tabular}{lr}
\toprule
Item & Result \\
\midrule
Within-set $Q$ correlation & -0.320285 \\
Within-set $\omega$ correlation & -0.405782 \\
Overlap with baseline model list & $11/265=4.1509\%$ \\
Share with strictly positive $G_i$ & 95.3578\% \\
Households with nonpositive $G_i$ & 144 \\
Minimum $G_i$ & -2.759967 \\
Parameters at bounds & 6/9 \\
Projected gradient (average loss) & $6.382625\times10^{-10}$ \\
Prespecified starts near lower loss & 9/20 \\
Lower solution: total objective & 364.748007 \\
Other valid solution: total objective & 365.527623 \\
\bottomrule
\end{tabular}
\end{table}

Multiple starting values approach the same lower objective, distinguishing this result from the failure of a single optimization run. Repeated starting values do not constitute independent statistical evidence or establish global optimality. Another solution is reached from starting values that include the specification's full-sample estimate; its projected-gradient norm on the average-loss scale is approximately $8.3\times10^{-7}$. It satisfies the numerical standard but has a higher objective value. The higher-loss solution is not substituted to preserve the original coefficient directions.

Under the same vector of multipliers, the original-weight softmax specification retains all three directions, has a within-set $Q$ correlation of 0.899673, and overlaps with the baseline model list by 70.1887\%. The equal-weight-across-sets pairwise specification also reverses the elderly-member direction, with a $Q$ correlation of 0.725092 and a list overlap of 43.7736\%. The interaction between the criterion and county reweighting reveals sensitivity in part of the perturbation distribution. Because 144 households fail strict positivity, the sharply different softmax solution lies outside the region in which the complete allocation score responds positively to the household component for every household; this condition is distinct from numerical convergence. The protocol does not add a positivity-constrained reestimation, introduce further temperatures, or exclude this perturbation. The 30 perturbations are not used to estimate the population probability of such an event.

\setcounter{table}{0}
\renewcommand{\thetable}{H.\arabic{table}}
\setcounter{figure}{0}
\renewcommand{\thefigure}{H.\arabic{figure}}

\section{Evidence Sources}
\label{app:evidence-sources}

The paper's main empirical results come from distinct stages of the original analysis. The household-level five-fold cross-fitted procedure supplies the food-consumption and education-spending gain inputs used by the structural model. The baseline structural estimation supplies the point estimates and their original numerical checks. The profile-search analysis supplies the parameter profiles, household-component signs, strict-positivity results, and fixed-quota allocation results. Table 1 uses a separately implemented county-held-out prediction diagnostic that does not exactly reproduce the original gain-generation workflow and therefore is not a household-level out-of-county validation of the generated gains used by the model. The comparison of five full-sample fitting criteria and the 30 county reweightings follows a prespecified protocol; the associated full-sample point estimates, 150 reweighted solutions, and optimization-start records are retained in the research archive. These analyses examine structural fit, parameter profiles, predictive performance, and allocation sensitivity as distinct objects rather than as a single test of parameter identification.

The remaining auxiliary analyses have different evidence sources and scopes. Appendix I reports supplementary logit and linear-probability entry benchmarks and their county-held-out evaluation; these models do not estimate a structural household component or government welfare weights. Appendix J uses the archived gain vectors and the structural sample to examine exact linear dependence between the generated gains and the seven household characteristics, without reestimating the first stage. Appendix K reports an additional conditional structural-model comparison whose county-held-out analysis continues to use the original fixed gains rather than regenerating them within each outer training fold. Appendix L reconstructs the list frontier from the baseline estimate and stored profile-search parameters without new structural-parameter estimation. The four analyses respectively provide nonstructural comparisons, examine the information structure of generated inputs, compare conditional model fit, and describe sensitivity among evaluated lists. They define the evidentiary scope of the main conclusions but do not constitute new independent sources of identification.

The parameter-substitution comparison in Section 6.2 uses the 57 strictly positive solutions in the specified 1\% raw-loss-tolerance region from Appendix E, with the fixed baseline estimate as the reference. For each parameterization, ordinary Spearman correlations of the complete log scores are first calculated within each county-by-urban/rural comparison set. Sets with finite correlations are retained, the correlations are averaged using the number of all household pairs in each set as weights, and the median is then taken across the 57 solutions. Holding the baseline household component fixed and replacing only the gain parameters gives a median correlation of 0.853. Using the corresponding conditionally reoptimized household component gives a median of 0.960. These within-set correlations do not use survey weights, and their pair-count weighting across sets differs from the equal-set weighting in Appendix G.

Across the same 57 strictly positive solutions, the within-set weighted covariance between changes in the log household component and changes in the log gain multiplier is negative in every case. The median descriptive variance-offset ratio is approximately 56\%.

\setcounter{table}{0}
\renewcommand{\thetable}{I.\arabic{table}}
\setcounter{figure}{0}
\renewcommand{\thefigure}{I.\arabic{figure}}

\section{Supplementary Simple Entry Benchmarks}
\label{app:simple-entry-benchmarks}

These supplementary benchmarks were implemented after the main structural and profile analyses and hold fixed the structural sample, definitions of the seven variables, survey weights, and geographic units. Logit and linear probability models (LPMs) fit observed entry among households that did not receive Dibao at baseline. A weighted comparison-set fixed-effects LPM provides an additional associational comparison. These models do not estimate the probability of approval conditional on application, welfare weights, or a structural household component.

\begin{table}[htbp]
\centering
\caption{Supplementary entry regressions}
\label{tab:app-entry-regressions}
\small
\begin{tabular}{lccc}
\toprule
Covariate & Logit & LPM & Set FE-LPM \\
\midrule
Income rank & -2.2672 & -0.1617 & -0.1644 \\
 & (0.4136) & (0.0265) & (0.0327) \\
Household size/10 & 1.9269 & 0.1453 & 0.1607 \\
 & (0.7276) & (0.0672) & (0.0688) \\
Child present & -0.2255 & -0.0128 & -0.0210 \\
 & (0.2719) & (0.0205) & (0.0204) \\
Elderly member present & 0.2935 & 0.0224 & 0.0203 \\
 & (0.2341) & (0.0174) & (0.0190) \\
Female household head & 0.3987 & 0.0280 & 0.0149 \\
 & (0.2352) & (0.0200) & (0.0189) \\
Age of household head/100 & 0.2951 & 0.0193 & -0.0018 \\
 & (1.0943) & (0.0856) & (0.0935) \\
Education code/9 & -0.8462 & -0.0364 & -0.0854 \\
 & (0.8970) & (0.0496) & (0.0547) \\
\bottomrule
\end{tabular}
\parbox{0.96\textwidth}{\footnotesize\textit{Note:} All specifications use the same 3,102 households, 265 entrants, 107 counties, and 125 comparison sets and include all seven covariates. Survey weights are normalized globally to have mean one. County-clustered standard errors are in parentheses. The finite-sample correction for the FE-LPM accounts for all 125 absorbed intercepts. $p$-values use a $t$ distribution with 106 degrees of freedom. The pooled regressions include an intercept; the FE-LPM absorbs comparison-set intercepts. The table is not used to compare the statistical significance of structural coefficients.}
\end{table}

The three focal directions are the same across these regressions. In the seven-variable logit, county-clustered $p$-values are below 0.001 for income, 0.348 for household-head education, and 0.213 for elderly-member status. The $p$-value for household size is 0.0093, while those for the remaining logit covariates exceed 0.05. The recurrence of signs in the structural sensitivity analysis is not a significance test.

All regression coefficients for a held-out fold are estimated using training counties only, and the five folds follow the archived common county split. Evaluation remains conditional on the prespecified common-support sample and ranking transformations; the full feature-construction process is not reestimated separately by fold. Table~\ref{tab:app-entry-prediction} pools held-out observations using survey weights. The small differences between the logit specifications are descriptive and do not establish statistically significant predictive superiority.

\begin{table}[htbp]
\centering
\caption{Supplementary county-held-out entry-prediction metrics}
\label{tab:app-entry-prediction}
\small
\begin{tabular}{lrrrr}
\toprule
Model/covariates & AUC & Log loss & Brier score & Out-of-range households \\
\midrule
Logit/income only & 0.6706 & 0.2774 & 0.07567 & 0 \\
Logit/seven variables & 0.6660 & 0.2787 & 0.07593 & 0 \\
LPM/income only & 0.6713 & 0.2791 & 0.07588 & 6 \\
LPM/seven variables & 0.6626 & 0.2895 & 0.07624 & 188 \\
\bottomrule
\end{tabular}
\parbox{0.96\textwidth}{\footnotesize\textit{Note:} Metrics are computed by pooling held-out predictions from all five folds with survey weights, not by taking an unweighted average of fold-level metrics. Features and the analysis sample remain fixed. AUC and Brier scores for the LPM use raw scores; log loss clips scores to $[10^{-10},1-10^{-10}]$. Out-of-range households are those with raw predictions outside $[0,1]$. No uncertainty test is conducted for differences across models.}
\end{table}

The seven-variable LPM produces 180 in-sample and 188 held-out predictions outside $[0,1]$, equal to 5.80\% and 6.06\% of the sample; all are below zero. Clipping predictions to $[0,1]$ changes its held-out Brier loss from 0.0762391 to 0.0761949 and its AUC from 0.6625722 to 0.6625144. LPM log loss therefore depends on the clipping convention. Fixed-quota comparisons use raw scores, retain the observed number of entrants in each comparison set, and divide the intersection between selected households and observed entrants by 265. Held-out list overlap is 24.15\% for the income-only logit and 26.79\% for the seven-variable logit; the corresponding LPM overlaps are 24.15\% and 27.17\%. These are conditional list comparisons, not overall classification-accuracy measures or ex ante predictions of quotas.

\setcounter{table}{0}
\renewcommand{\thetable}{J.\arabic{table}}
\setcounter{figure}{0}
\renewcommand{\thefigure}{J.\arabic{figure}}

\section{Exact Dependence of the Implemented Gain Predictions}
\label{app:gain-dependence}

The allocation score combines the household component with predicted policy gains. Distinguishing their relative contributions therefore requires examining the information on which each component depends and the parameter restrictions imposed by the model. In the general BBK framework, separation is aided when some variables that affect predicted policy gains do not enter the household component directly. Even among households with similar baseline allocation conditions, the additional variation in predicted gains can then further restrict the outcome-gain coefficients. When both components depend on the same household characteristics, parameter separation relies more heavily on the specific functional form of the allocation score and on model restrictions. This information structure shapes the empirical interpretation of the parameters and makes substitution and compensation between the household component and predicted gains an object of analysis.

In an additive logit or linear probability model containing an intercept and the seven household characteristics, the two implemented gain predictions are perfectly collinear with the existing covariates, so their separate coefficients cannot be estimated. Parameter separation in the multiplicative score instead depends on its nonlinear mapping and model restrictions; Section 3.3 states the relevant information conditions (Tamer, 2010).

The implemented R-learner estimates a weighted regression of residuals, with an intercept and the seven household characteristics in its heterogeneity function. Its prediction function is therefore affine in those characteristics. After the original gain predictions are joined to the structural analysis frame by the household-window identifier, the differences in all seven characteristic columns and both gain columns are zero within the structural sample. This check uses the archived vectors and does not reestimate the first stage.

The eight-column matrix $[1,X]$ has rank 8. After the two gain predictions are added, the resulting ten-column matrix remains rank 8 under absolute tolerances from $10^{-12}$ to $10^{-8}$. Its two smallest singular values are approximately $1.37\times10^{-14}$ and $8.72\times10^{-15}$, with a default tolerance of $5.86\times10^{-11}$. The maximum absolute least-squares reconstruction error is $3.16\times10^{-15}$ for food and $7.99\times10^{-15}$ for education. We therefore do not interpret separate coefficients on additional additive gain regressors.

These calculations establish dependence among the generated inputs in this sample; they do not imply an absence of heterogeneity in the true treatment effects. The nonlinear restrictions imposed by the multiplicative exponential model can support numerical fit, but they do not supply the independent variation needed to interpret the household component and gain parameters as separate economic primitives.

\setcounter{table}{0}
\renewcommand{\thetable}{K.\arabic{table}}
\setcounter{figure}{0}
\renewcommand{\thefigure}{K.\arabic{figure}}

\section{Conditional Structural-Model Comparison}
\label{app:conditional-model-comparison}

A supplementary conditional exercise compares a linear score, an exponential household score without gains, and the complete multiplicative score, all using the same entrant-weighted softmax objective. The seven score coefficients share the bounds $[\log(0.05),4]$; the complete model adds two gain coefficients bounded by $[0,10]$. None of the models includes an intercept or an added positivity constraint. Identical numerical bounds constrain the three functional forms differently and do not give them equal flexibility.

\begin{table}[htbp]
\centering
\caption{Supplementary conditional structural exercise}
\label{tab:app-conditional-structural}
\small
\begin{tabular}{lrrrr}
\toprule
Score & Full-sample loss & Held-out loss & Full-sample overlap & Held-out overlap \\
\midrule
Linear score & 765.133 & 783.607 & 75/265 & 69/265 \\
Exponential score without gains & 765.246 & 776.152 & 79/265 & 72/265 \\
Complete BBK score & 763.514 & 785.236 & 77/265 & 76/265 \\
\bottomrule
\end{tabular}
\parbox{0.96\textwidth}{\footnotesize\textit{Note:} All models use the same weighted softmax criterion, temperature, sample, comparison sets, and county folds. Held-out loss is the sum of the five fold losses and retains globally normalized weights. Overlap is the intersection with observed entrants after households are selected using the observed quota in each comparison set, with 265 as the denominator. The original gain predictions are not reestimated within the outer training folds. This is a conditional held-out comparison given the fixed generated gains, not a strict out-of-county evaluation of the complete BBK workflow.}
\end{table}

The full-sample fits use seven starting values and at most 950 iterations; each training-fold fit uses four starting values and at most 650 iterations, with fixed seeds and common L-BFGS-B tolerances. Every selected fit reports successful optimization. These checks support numerical comparability within this exercise but do not establish global optimality. The full-sample losses reproduce the supplied results at the reported precision. The complete model has lower in-sample loss but higher conditional held-out loss than the two simpler scores. No uncertainty analysis is conducted to establish a statistical ranking among the models.

The complete model always uses the original household-level cross-fitted gain predictions, which were not generated by a first-stage procedure that fully excluded each test county. The results are therefore conditional held-out comparisons given the fixed generated gains. They do not establish that gains are irrelevant, that administrative decisions ignore outcomes, or that the complete BBK workflow has better or worse out-of-county predictive performance. This auxiliary comparison defines the information on which gain-related interpretations rest.

\setcounter{table}{0}
\renewcommand{\thetable}{L.\arabic{table}}
\setcounter{figure}{0}
\renewcommand{\thefigure}{L.\arabic{figure}}

\section{Construction of the Evaluated List Frontier}
\label{app:list-frontier}

This extension uses the archived structural analysis frame, the baseline estimate, and all 1,768 stored profile-search parameterizations; no parameters are reestimated. At each point, we calculate $G_i=1+\beta_C\widehat\tau_{iC}+\beta_E\widehat\tau_{iE}$ and $Q_i=\exp(X_i'\theta)G_i$, break ties stably using the original household key, and select the top $K_m$ households in each comparison set. The main-box curve uses the 1,752 points for which both gain coefficients lie in $[0,10]$; the other 16 extended points remain in the summaries for the full evaluated grid. The historical 5\% band contains 587 main-box points, whose lowest overlap with the baseline list is 21.5\%. The corresponding value across the complete set of 603 points is 12.8\%.

\begin{figure}[htbp]
\centering
\includegraphics[width=\textwidth]{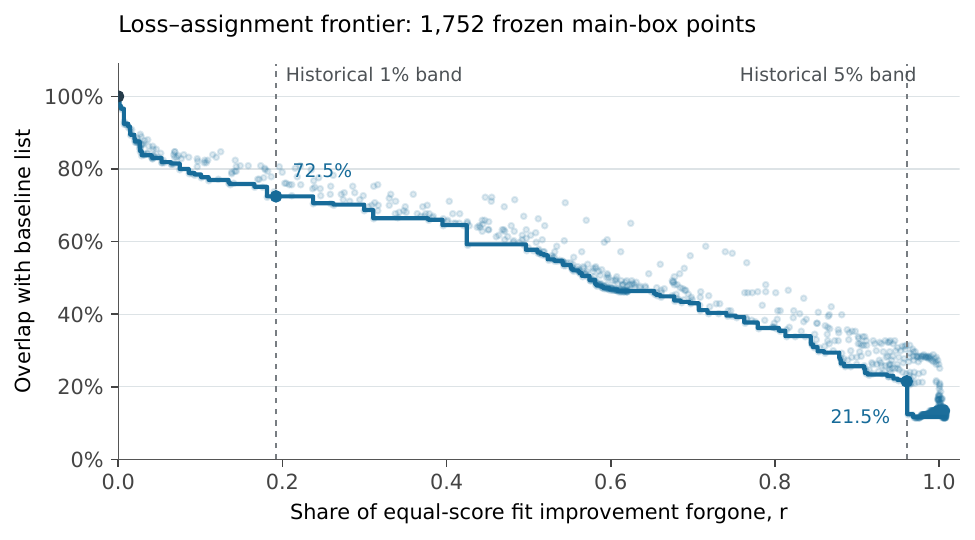}
\caption{Loss and list frontier over the evaluated main parameter box}
\label{fig:app-loss-list-frontier}
\parbox{0.96\textwidth}{\footnotesize\textit{Note:} The figure uses the 1,752 stored profile solutions inside the main parameter box together with the baseline solution. The step function gives the lowest overlap with the baseline model list among evaluated solutions within each normalized loss tolerance. Reference lines correspond to the historical 1\% and 5\% raw-total-loss bands. The envelope is formed only over evaluated solutions, without interpolation or parameter reestimation.}
\end{figure}

Within the main parameter box, the analysis uses the 1,752 stored profile solutions and the baseline estimate as the reference. For each loss tolerance $t$, we retain all evaluated solutions satisfying $r(L_s)\leq t$, generate simulated lists using the fixed-quota rule in Section 3.5, calculate overlap with the baseline model list, and record the minimum overlap in that set. This procedure produces the step frontier in Figure~\ref{fig:app-loss-list-frontier}. Every list preserves the observed entrant count in each comparison set, uses the original household identifier to break score ties, and uses the 265 simulated positions as the denominator for overlap. The frontier links acceptable model fit to household selection by reporting the largest list deviation supported by the evaluated parameter search at each fit requirement.

When $r(L)\leq0.05$, so that a parameter combination retains at least 95\% of the baseline model's fit improvement over the equal-score benchmark, every eligible evaluated solution overlaps with the baseline list by at least 83.0\%. The eligible lists therefore remain similar under this stricter fit requirement even though the structural parameters can still change. List differences expand as the loss tolerance increases. At $r(L)=0.961$, which corresponds to the historical 5\% raw-total-loss band, the minimum baseline-list overlap within the main parameter box falls to 21.5\%. When all evaluated solutions in the main box are included, the minimum falls further to 11.7\%. The small common-selection core reported in Appendix F thus occurs over a broad loss-tolerance range. Evaluated parameterizations constrain household selection much more strongly when they must preserve most of the in-sample fit improvement; substantially different lists emerge only when larger fit losses are permitted.

The frontier above uses stored solutions within the main parameter box and therefore has a different comparison set from the 603 solutions satisfying the specified 5\% raw-loss-tolerance criterion in Appendix F, which include extended coordinates. Within the historical 5\% raw-total-loss band, the main box contains 587 evaluated parameterizations and has a minimum baseline-list overlap of 21.5\%. Adding the extended coordinates outside the main box yields the complete 603-point set and lowers the minimum overlap to 12.8\% (Table~\ref{tab:app-evaluated-sets}). List stability therefore depends on both the loss tolerance and the parameter-search range. Parameter combinations that preserve more of the in-sample fit improvement tend to support lists closer to the baseline result, while broader loss tolerances or parameter ranges support more varied household selections.

Let $A_0$ denote the baseline model list and $A_s$ the list generated by stored solution $s$. Baseline overlap is $|A_s\cap A_0|/265$, pairwise overlap is $|A_s\cap A_u|/265$, and common retention is $|A_0\cap(\bigcap_s A_s)|/265$. Always selected, never selected, and status-changing households are classified over all 3,102 households. The strict-positivity subset requires $G_i>0$ for every household and contains 57 points. The strict-positivity column in Table~\ref{tab:app-evaluated-sets} uses only these 57 points. Figure 2 in the main text additionally includes the positive baseline solution as an $r=0$ anchor; adding that anchor does not change any value reported in the table.

For each tolerance $t$, the baseline-overlap frontier takes the lowest overlap among recorded solutions with $r_s\leq t$. The upper and lower envelopes for group composition similarly take the minimum and maximum list shares. Income quintiles are equal-sized groups formed from the income-rank variable in the structural sample, and elderly-member status uses the original binary indicator. The curves are step functions over evaluated solutions and remain conditional on the model, loss, baseline reference, parameter box, and grid. They do not cover other unrecorded low-loss household-component vectors.

If a household's baseline rank within its comparison set is $h=1,\ldots,N_m$, its distance from the cutoff pair is $d=\min\{|h-K_m|,|h-(K_m+1)|\}$. A status-change event occurs when a household's selection under a stored solution differs from its baseline status; the same household can generate events under multiple solutions. The 57-, 103-, and 603-point sets produce 4,712, 9,006, and 146,216 events, respectively. The shares within one rank of the cutoff and at least three ranks away describe the location of these events and do not identify a causal role for score gaps. Because county and urban/rural status jointly define comparison sets and each set's quota is fixed, these comparisons do not change the total rural selection quota.

\begin{table}[htbp]
\centering
\caption{Allocation objects across evaluated parameter sets}
\label{tab:app-evaluated-sets}
\footnotesize
\resizebox{\textwidth}{!}{%
\begin{tabular}{lccc}
\toprule
Metric & Strict positivity: 57 points & Historical 1\% band: 103 points & Historical 5\% band: 603 points \\
\midrule
Minimum baseline-list overlap & 72.5\% & 72.5\% & 12.8\% \\
Minimum pairwise list overlap & 66.8\% & 60.0\% & 3.0\% \\
Baseline-list households retained at every point & 58.1\% & 44.5\% & 1.1\% \\
Always selected / never selected / status changes & 154 / 2,698 / 250 & 118 / 2,625 / 359 & 3 / 1,860 / 1,239 \\
Share of changes within one rank of cutoff & 86.3\% & 85.4\% & 56.7\% \\
Share of changes at least three ranks from cutoff & 6.1\% & 7.0\% & 31.4\% \\
Lowest-income-quintile share of list & 72.5\%--80.4\% & 72.5\%--80.4\% & 38.5\%--80.4\% \\
Households with elderly members: share of list & 78.1\%--88.3\% & 77.4\%--88.3\% & 2.6\%--97.4\% \\
\bottomrule
\end{tabular}%
}
\parbox{0.96\textwidth}{\footnotesize\textit{Note:} The strict-positivity set contains the 57 evaluated profile solutions satisfying $G_i(\beta)>0$ for all 3,102 households. The 103- and 603-point sets satisfy the historical 1\% and 5\% raw-loss-tolerance criteria; the 603-point set includes extended coordinates outside the main parameter box. The baseline solution is not added to any of the three sets in the table. List-overlap rates and group shares use the 265 simulated positions as the denominator, while household-status classifications use all 3,102 households. Cutoff-distance statistics are defined over household-by-parameterization status-change events, so a household can be counted more than once. The results describe the evaluated parameter sets; they are not confidence intervals or household selection probabilities.}
\end{table}

\end{document}